\documentclass[sigconf,nonacm]{acmart}
\usepackage{enumitem}
\usepackage{xspace}
\usepackage{array}
\usepackage{adjustbox}
\usepackage{stfloats}   
\usepackage{placeins}   
\usepackage{algorithm}
\usepackage{algpseudocode}
\usepackage{tikz}
\usetikzlibrary{arrows.meta,positioning,decorations.pathreplacing,calc}
\usepackage{pgfplots}
\pgfplotsset{compat=1.17}

\definecolor{secfill}{HTML}{FBE1C8}
\definecolor{secline}{HTML}{C8691C}
\definecolor{pubfill}{HTML}{D9E6F4}
\definecolor{publine}{HTML}{3B73AE}
\definecolor{locfill}{HTML}{E3F1DF}
\definecolor{locline}{HTML}{3F7A35}
\definecolor{comfill}{HTML}{F8E0DC}
\definecolor{comline}{HTML}{A8372A}
\definecolor{srvgreen}{HTML}{6A7F68}
\definecolor{srvbrown}{HTML}{9A5B45}
\definecolor{mpcline}{HTML}{5B2A5E}
\definecolor{mpcfill}{HTML}{FBF8FC}

\newcommand{\model}{SlimDTA\xspace}
\newcommand{\ring}{\ensuremath{\mathbb{Z}_{2^{\ell}}}\xspace}
\newcommand{\share}[1]{\ensuremath{[\![ #1 ]\!]}}
\newcommand{\msb}{\ensuremath{\mathrm{MSB}}}

\newcommand{\relu}{\ensuremath{\mathrm{ReLU}}}
\newcommand{\ind}[1]{\ensuremath{\mathbf{1}[#1]}}
\newcommand{\xlow}{\ensuremath{\hat x_{\mathrm{low}}}}
\newcommand{\rlow}{\ensuremath{r_{\mathrm{low}}}}

\begin{document}

\title[GPU-Accelerated Secure Inference for Drug--Target Interaction Prediction]{GPU-Accelerated Secure Inference for \\ Drug--Target Interaction Prediction}

\author{Sikha Pentyala}
\affiliation{%
  \institution{University of Washington Tacoma}
  \city{Tacoma}
  \state{Washington}
  \country{USA}}
\author{Steven Golob}
\affiliation{%
  \institution{University of Washington Tacoma}
  \city{Tacoma}
  \state{Washington}
  \country{USA}}
\author{Daniil Filienko}
\affiliation{%
  \institution{University of Washington Tacoma}
  \city{Tacoma}
  \state{Washington}
  \country{USA}}
\author{Ming-Chun Kao}
\affiliation{%
  \institution{University of Washington Tacoma}
  \city{Tacoma}
  \state{Washington}
  \country{USA}}
\author{Martine De Cock}
\affiliation{%
  \institution{University of Washington Tacoma}
  \city{Tacoma}
  \state{Washington}
  \country{USA}}
\renewcommand{\shortauthors}{Pentyala et al.}

\begin{abstract}
Deep learning models that predict drug--target interactions (DTI) can speed up drug discovery, but the candidate drugs they score are valuable intellectual property that their owners cannot reveal. Secure multiparty computation (MPC) lets two servers compute a prediction on a secret-shared drug, without either server learning the drug. However, secure inference of deep models is slow. Non-linear operations require communication between the servers, and MPC arithmetic does not map naturally to GPUs. In this paper, we present our solution to Track~III of the iDASH 2026 privacy and security competition, which asked for fast two-party secure DTI inference on GPUs. Our solution is built on the co-design of MPC-friendly models and MPC protocols. On the model side, we design \model, a DTI model that reduces the operations on the secret drug, and in particular the non-linear operations that require communication. On the protocol side, we build a GPU framework that evaluates the linear layers with additive secret sharing and the non-linear layers with function secret sharing, including a custom CUDA kernel for the comparisons. Our solution won first place in the competition, at under 10\,ms per prediction on average, and its secure predictions match the plaintext model to within 0.02\%. 
\end{abstract}


\maketitle

\section{Introduction}

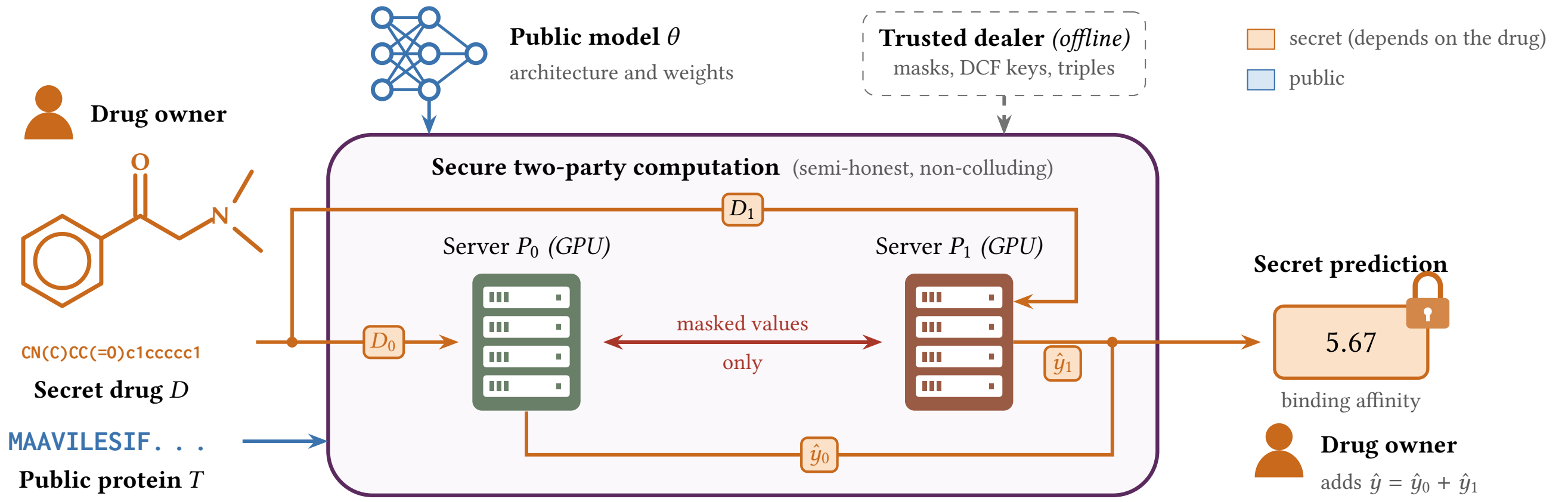
\begin{figure*}[b]
\centering
\begin{adjustbox}{max width=\textwidth}
\begin{tikzpicture}[font=\small, >={Stealth[length=2.2mm]},
  lab/.style={font=\small, align=center},
  sub/.style={font=\footnotesize, align=center, text=black!65},
  chip/.style={draw=secline, fill=secfill, rounded corners=2pt, font=\footnotesize, inner sep=2pt, line width=0.8pt},
  sarr/.style={->, line width=1.1pt, secline},
  srvA/.style={fill=srvgreen},
  srvB/.style={fill=srvbrown},
  server/.pic={
    \fill[rounded corners=3pt, pic actions] (-0.75,-0.95) rectangle (0.75,0.95);
    \foreach \y in {0.62,0.21,-0.20,-0.61}{
      \fill[white, rounded corners=1pt] (-0.6,\y-0.15) rectangle (0.6,\y+0.15);
      \foreach \x in {-0.48,-0.38,-0.28}
        \fill[pic actions] (\x-0.03,\y-0.07) rectangle (\x+0.03,\y+0.07);
      \fill[pic actions] (0.42,\y-0.04) rectangle (0.48,\y+0.04);
    }
  },
  person/.pic={
    \fill[secline] (0,0.2) circle (0.15);
    \fill[secline] (-0.27,-0.25) arc (180:0:0.27 and 0.3) -- cycle;
  },
  molecule/.pic={
    \begin{scope}[draw=secline, line width=1.3pt, line cap=round, line join=round]
      \draw (30:0.5) \foreach \a in {90,150,...,390} { -- (\a:0.5) };
      \draw (0,0) circle (0.3);
      \draw (30:0.5) -- ++(30:0.5) coordinate (c);
      \draw ([xshift=-1.6pt]c) -- ++(0,0.45);
      \draw ([xshift=1.6pt]c) -- ++(0,0.45);
      \node[text=secline, font=\small\sffamily\bfseries, inner sep=1pt] at ($(c)+(0,0.6)$) {O};
      \draw (c) -- ++(-30:0.5) coordinate (d) -- ++(30:0.42) coordinate (nn);
      \node[text=secline, font=\small\sffamily\bfseries, inner sep=1pt] (N) at ($(nn)+(30:0.1)$) {N};
      \draw (N.north east) -- ++(60:0.38);
      \draw (N.south east) -- ++(-40:0.38);
    \end{scope}
  },
  network/.pic={
    \foreach \y [count=\i] in {0.5,0,-0.5} \coordinate (in\i) at (0,\y);
    \foreach \y [count=\i] in {0.5,0,-0.5} \coordinate (hd\i) at (0.65,\y);
    \coordinate (out) at (1.3,0);
    \foreach \i in {1,2,3}{\foreach \j in {1,2,3} \draw[publine, line width=0.9pt] (in\i) -- (hd\j); \draw[publine, line width=0.9pt] (hd\i) -- (out);}
    \foreach \p in {in1,in2,in3,hd1,hd2,hd3,out} \draw[publine, line width=1.3pt, fill=white] (\p) circle (0.13);
  },
  padlock/.pic={
    \draw[secline, line width=1.6pt] (-0.14,0.1) -- (-0.14,0.25) arc (180:0:0.14) -- (0.14,0.1);
    \fill[secline!80, rounded corners=1.5pt] (-0.24,-0.22) rectangle (0.24,0.12);
    \fill[white] (0,-0.03) circle (0.045);
    \fill[white] (-0.018,-0.03) rectangle (0.018,-0.14);
  }]

\draw[rounded corners=14pt, line width=1.2pt, draw=mpcline, fill=mpcfill] (3.4,-1.45) rectangle (12.6,2.55);
\node at (8.0,2.17) {\bfseries Secure two-party computation \ {\footnotesize\mdseries\color{black!65}(semi-honest, non-colluding)}};
\pic[srvA, scale=0.8] at (5.6,0.25) {server};
\pic[srvB, scale=0.8] at (10.4,0.25) {server};
\node[lab] at (5.6,1.3) {Server $P_0$ \textit{(GPU)}};
\node[lab] at (10.4,1.3) {Server $P_1$ \textit{(GPU)}};
\draw[<->, line width=1.2pt, comline] (6.45,0.25) -- node[above, sub, text=comline] {masked values} node[below, sub, text=comline] {only} (9.55,0.25);

\pic at (0.3,2.75) {person};
\node[font=\small\bfseries, anchor=west] at (0.65,2.75) {Drug owner};
\pic at (0.45,1.15) {molecule};
\node[font=\scriptsize\ttfamily\bfseries, text=secline] at (1.0,0.12) {CN(C)CC(=O)c1ccccc1};
\node[lab] at (1.0,-0.3) {\textbf{Secret drug} $D$};

\draw[line width=1.1pt, secline] (2.6,0.25) -- (3.0,0.25);
\fill[secline] (3.0,0.25) circle (0.06);
\draw[sarr] (3.0,0.25) -- node[pos=0.55, chip] {$D_0$} (4.85,0.25);
\draw[sarr] (3.0,0.25) |- (11.7,1.72) |- (11.0,0.7);
\node[chip] at (8.0,1.72) {$D_1$};

\node[font=\normalsize\ttfamily\bfseries, text=publine] at (1.0,-0.85) {MAAVILESIF\ldots};
\node[lab] at (1.0,-1.3) {\textbf{Public protein} $T$};
\draw[->, line width=1.1pt, publine] (2.45,-0.85) -- (3.4,-0.85);

\pic[scale=0.8] at (4.0,3.45) {network};
\node[lab, anchor=west] at (5.3,3.65) {\textbf{Public model} $\theta$};
\node[sub, anchor=west] at (5.3,3.22) {architecture and weights};
\draw[->, line width=1.1pt, publine] (4.52,2.95) -- (4.52,2.55);

\node[draw=black!55, dashed, rounded corners=4pt, fill=white, align=center, inner sep=5pt] (deal) at (10.9,3.45)
  {\textbf{Trusted dealer} \textit{(offline)}\\[-1pt] {\footnotesize\color{black!65} masks, DCF keys, triples}};
\draw[->, line width=1.1pt, dashed, black!55] (deal.south) -- (10.9,2.55);

\draw[line width=1.1pt, secline] (5.6,-0.52) -- (5.6,-1.0) -- node[pos=0.5, chip] {$\hat y_0$} (12.1,-1.0) -- (12.1,0.25);
\draw[line width=1.1pt, secline] (11.0,0.25) -- node[below=1pt, chip] {$\hat y_1$} (12.1,0.25);
\fill[secline] (12.1,0.25) circle (0.06);
\draw[sarr] (12.1,0.25) -- (13.75,0.25);
\node[lab] at (14.75,1.1) {\textbf{Secret prediction}};
\node[draw=secline, fill=secfill, line width=1pt, rounded corners=3pt, minimum width=1.7cm, minimum height=0.8cm, font=\large] (y) at (14.75,0.25) {$5.67$};
\pic at (15.6,0.62) {padlock};
\node[sub] at (14.75,-0.42) {binding affinity};
\pic at (13.95,-1.0) {person};
\node[font=\small\bfseries, anchor=west] at (14.3,-0.92) {Drug owner};
\node[sub, anchor=west] at (14.3,-1.32) {adds $\hat y = \hat y_0 + \hat y_1$};

\fill[secfill, draw=secline] (13.6,3.5) rectangle ++(0.3,0.22);
\node[sub, anchor=west] at (13.95,3.61) {secret (depends on the drug)};
\fill[pubfill, draw=publine] (13.6,3.05) rectangle ++(0.3,0.22);
\node[sub, anchor=west] at (13.95,3.16) {public};
\end{tikzpicture}
\end{adjustbox}
\caption{Secure DTI inference as a service. The drug owner splits its secret
drug into two shares and sends one to each server. The protein and the model
are public, and both servers know them. The servers run the secure protocol on
their GPUs and exchange only masked values, which look uniformly random. Each
server returns its share of the prediction, and only the drug owner can add the
two shares to obtain $\hat y$. A trusted dealer prepares masks, keys, and
triples offline.}
\label{fig:setting}
\end{figure*}

Predicting drug--target interactions (DTI) is a core step in computational drug discovery. A DTI model takes a candidate drug and a target protein, and predicts how strongly the two bind. These predictions help researchers narrow a large chemical space down to a few promising molecules, repurpose existing drugs, and anticipate off-target effects, all before running costly laboratory experiments. Deep learning models, from convolutional networks such as DeepDTA~\citep{ozturk2018deepdta} to Transformer-based models~\citep{huang2021moltrans} and multitask models such as DeepDTAGen~\citep{shah2025deepdtagen}, have steadily improved the accuracy of these predictions on standard benchmarks.

In practice, the drug and the model can be held by different parties. For example, a pharmaceutical company may want to screen its new drug candidates against a target protein, using a DTI model developed by another organization. The chemical structures of the candidates are the company's intellectual property, and the model may be proprietary as well. Such a collaboration requires one party to trust the other with its data or model.

Secure multiparty computation (MPC) removes the need for this trust. In MPC, the drug and the model are split into secret shares and distributed among non-colluding servers. The servers compute the prediction on the shares, without any server learning the drug, the model, or any intermediate value. \citet{hie2018realizing} showed that MPC can enable such pharmacological collaboration with conventional machine learning models.

The concern is efficiency. Modern DTI models are deep neural networks, and evaluating them in MPC is far more expensive than in the clear. In MPC, additions and multiplications by public values are cheap, but multiplying two secret values and computing non-linear functions, such as activations and comparisons, require the servers to communicate (Section~\ref{sec:primer}). A deep model performs many such operations for every prediction. Moreover, most MPC frameworks were designed for CPUs. GPUs, the hardware that made deep learning practical, are optimized for floating-point matrix operations, not for the modular integer arithmetic that MPC relies on.\footnote{MPC protocols compute over the integers modulo $2^{\ell}$, typically with $\ell = 32$ or $\ell = 64$. We use $\ell = 64$, which leaves room to accumulate large dot products in fixed point without overflow. GPU tensor cores accelerate matrix products of floating-point numbers and of 8-bit integers, but not of 32-bit or 64-bit integers.} As a result, secure inference of a modern DTI model is too slow to score large batches of drug candidates.

Track~III of the iDASH 2026 privacy and security competition\footnote{\url{https://humangenomeprivacy.org/2026/competition-tasks.html}} posed this question. Can a two-party MPC framework evaluate a deep DTI model on GPUs, with accuracy close to the reference DeepDTAGen model~\citep{shah2025deepdtagen} and with high efficiency? The challenge fixed a setting in which only the drug is confidential, while the target protein and the model weights are treated as public (Section~\ref{sec:setting}). This choice keeps the focus on the secure computation on the drug, which is the expensive part. It is also practical. Target proteins are often publicly known, and a model can be published or licensed, while the drug candidate is the asset that must be protected. Unlike the setting of \citet{hie2018realizing}, where both drugs and targets are secret, one input is secret and the other is public. With public weights, the linear layers become cheap, and the non-linear layers remain the main cost.

Motivated by this challenge, we developed a solution built on one idea, the \emph{co-design of MPC-friendly models and MPC protocols}. Most systems for secure inference take the model as given and make the protocol faster (Section~\ref{sec:related}). Most DTI models are designed for accuracy in the clear, without regard to the cost of secure computation. We do both together. We shape the model to the costs of the protocol, and we shape the protocol and its GPU implementation to the operations that the model needs. This main contribution rests on two specific contributions.

\begin{itemize}[leftmargin=1.4em,itemsep=2pt]
\item \textbf{MPC-friendly model design.} We propose \model, an MPC-friendly DTI model. It reduces the number of operations on the secret drug overall, and in particular the number of non-linear operations, which are the ones that require communication. It also uses the asymmetry of the setting. Its secret drug tower is small, while its public protein tower, which costs nothing under MPC, carries most of the model's capacity. \model is more than four times smaller than DeepDTAGen, without losing accuracy (Section~\ref{sec:model}).
\item \textbf{GPU acceleration of the MPC protocol.} We build a GPU framework for secure two-party inference. The linear layers, which make up most of the computation, run with additive secret sharing, locally on each server's GPU, as exact integer matrix products. The non-linear layers, the only ones that communicate, run with function secret sharing, which keeps their communication small, and with a custom CUDA kernel, which makes their local computation fast. The framework processes inputs in large batches, and splits large input sets into batches that fit in GPU memory, to support any number of inputs (Sections~\ref{sec:gpu} and~\ref{sec:secure}).
\end{itemize}

Our solution won first place in Track~III. In the competition's evaluation on two GPU workstations, it processed batches of drug candidates in under 10\,ms per prediction on average, in all three tests on the BindingDB and Davis datasets. Our own experiments show that both contributions are needed to reach this speed (Section~\ref{sec:results}). Our code is publicly available.\footnote{\url{https://github.com/sikhapentyala/mpc-framework-idash}} We discuss the limitations of our solution in Section~\ref{sec:discussion}.

\section{Problem Setting}
\label{sec:setting}

\paragraph{The prediction task.} A DTI model $M_\theta$ with weights $\theta$ takes a drug $D$ and a target protein $T$, and outputs a predicted binding affinity $\hat y = M_\theta(D, T)$. The drug is given as a SMILES string~\citep{weininger1988smiles}, a line of text that encodes the structure of a molecule, such as \texttt{CC(=O)Oc1ccccc1C(=O)O} for aspirin. The protein is given as its sequence of amino acids, a string over an alphabet of 25 letters, such as \texttt{MAAVILESIFLKRS}\ldots. The affinity is a real number, for example the negative logarithm of the dissociation constant ($pK_d$). A higher value means stronger binding. The competition uses two public benchmarks, Davis~\citep{davis2011comprehensive} and BindingDB~\citep{liu2007bindingdb}. We also use KIBA~\citep{tang2014making} to evaluate the accuracy of our model in plaintext. To report accuracy, the predicted affinity is turned into a binary decision (binds or does not bind) with a threshold, and we measure its balanced accuracy.

\paragraph{Data holders and computing parties.} The drug owner, for example a pharmaceutical company, holds the candidate drug $D$, which is confidential. The model owner holds the trained weights $\theta$, and the target protein $T$ usually comes from a public database. Following the competition, we treat $\theta$, $T$, and every intermediate value that depends only on $T$ as public. Every value that depends on the drug, including the final prediction, must stay secret. One may wonder why MPC is needed when only one input is confidential, since the drug owner could in principle run the public model itself. In practice, the drug owner may not have the model deployment or the GPU infrastructure, and wants to use a prediction service without revealing its drugs. The competition fixed this setting to focus on the cost of the secure computation on the drug, which is the expensive part.\footnote{Our framework also supports secret weights. In that case, the linear layers need Beaver triples (Section~\ref{sec:primer}), at a higher cost that we measure in Table~\ref{tab:ablation}.}

The computation is outsourced to two servers, $P_0$ and $P_1$, in the MPC-as-a-service model (Figure~\ref{fig:setting}). Each server has a GPU. The servers can be run by two independent cloud providers, or by the model owner and an independent organization. The drug owner splits $D$ into two secret shares, as we explain in Section~\ref{sec:primer}, and sends one share to each server. Both servers know $T$ and $\theta$. The servers run the secure protocol, exchanging messages with each other, and return their shares of the prediction $\hat y$ to the drug owner, who adds them to obtain $\hat y$. A trusted \emph{dealer} prepares correlated randomness for the servers, such as random masks and keys, offline and before the drug is known.

\paragraph{Threat model.} We follow the threat model that is standard in the secure inference literature. The servers are \emph{semi-honest}, also called honest-but-curious. Each server follows the protocol, but it may inspect its own shares and the messages it receives, and try to learn the drug from them. The servers are \emph{non-colluding}, meaning they do not combine what they see. The dealer is trusted to generate correct randomness and not to collude with either server, and it never sees any input. Under these assumptions, neither server learns anything about the drug beyond public information, such as the padded input length and the batch size.

\paragraph{Efficiency metrics.} The competition measures the \emph{online} time, the time from the moment the servers hold the shares of the drugs until they hold the shares of the predictions. The \emph{offline} phase, in which the dealer prepares randomness, does not depend on the input and is not timed. The number of inputs to be scored can range from hundreds to hundreds of thousands of samples. The secure predictions must stay within 2\% of the accuracy of the plaintext reference model, and we use cryptographic primitives that offer 128-bit security.

The online time consists of computation and communication. Communication costs time in two ways. Each \emph{round}, in which a server must wait for a message from the other server before it can continue, costs at least one network latency $\tau$. Each byte costs time according to the bandwidth $\mathrm{BW}$ of the link. For a protocol with $R$ rounds that sends $S$ bytes,
\begin{equation}
T_{\mathrm{online}} \;\approx\; T_{\mathrm{comp}} \;+\; R \cdot \tau
\;+\; \frac{S}{\mathrm{BW}} .
\label{eq:cost}
\end{equation}
For example, on a 1\,Gbps link, sending one megabyte takes 8\,ms. A GPU reduces $T_{\mathrm{comp}}$, a good protocol reduces $R$ and $S$, and a good model reduces the number of expensive operations in the first place.

\section{A Primer on Secure Two-Party Computation}
\label{sec:primer}

This section introduces the cryptographic tools that our solution is built on. For a broader introduction, please refer to \citet{evans2018pragmatic} and \citet{cramer2015secure}. Table~\ref{tab:notation} summarizes the notation.

\begin{table}[t]
\centering
\caption{Notation.}
\label{tab:notation}
\small
\begin{tabular}{@{}l>{\raggedright\arraybackslash}p{0.62\columnwidth}@{}}
\toprule
Symbol & Meaning \\
\midrule
$P_0, P_1$ & the two servers \\
$\ring$ & the ring of integers modulo $2^{\ell}$; we use $\ell = 64$ \\
$f$ & number of fractional bits in fixed-point encoding; we use $f = 16$ \\
$\share{x} = (x_0, x_1)$ & additive secret sharing of $x$; server $P_i$ holds $x_i$ \\
$\msb(x)$ & most significant bit of $x$; equals 1 if and only if $x$ is negative \\
$r$, $\hat x = x + r$ & a random mask and a masked value \\
$\xlow$, $\rlow$ & the lower $\ell - 1$ bits of $\hat x$ and of $r$ \\
$k_0, k_1$ & the two keys of a function secret sharing \\
$\lambda$ & security parameter, the length of PRG seeds; we use $\lambda = 128$ \\
\bottomrule
\end{tabular}
\end{table}

\subsection{Secure Multiparty Computation}

Secure multiparty computation (MPC) is an umbrella term for cryptographic techniques that let mutually distrustful parties compute a function of their private inputs, while revealing nothing beyond the output~\citep{yao1986generate, goldreich1987play}. We use MPC as a service with two servers, as described in Section~\ref{sec:setting}, a setting called two-party computation (2PC). MPC protocols are designed to stay secure even if an adversary corrupts some of the parties. A \emph{passive}, or semi-honest, adversary follows the protocol, but tries to learn private information from what it sees. An \emph{active}, or malicious, adversary may also deviate from the protocol in arbitrary ways. Protocols secure against active adversaries need additional checks, and are therefore more expensive. Like most work on secure inference, we consider passive adversaries (Section~\ref{sec:setting}).

Many MPC protocols, including ours, are built on \emph{secret sharing}. The idea is to split a private value into several \emph{shares}, using randomness, and to give one share to each server, such that no single share reveals anything about the value. The value can be reconstructed only by combining the shares. The servers then compute on their shares, without reconstructing the inputs or any intermediate value.

\subsection{Additive Secret Sharing}
\label{sec:ass}

In \emph{additive} secret sharing, the shares add up to the secret. We compute over the ring \ring, the integers modulo $2^{\ell}$. Working modulo $2^{\ell}$ is natural for computers, because a 64-bit processor already computes modulo $2^{64}$ when an addition or multiplication overflows~\citep{cramer2018spdz2k}. A value $x \in \ring$ is \emph{additively shared} by choosing $x_0$ uniformly at random from \ring and setting $x_1 = x - x_0 \bmod 2^{\ell}$. Then
\begin{equation}
x_0 + x_1 \equiv x \pmod{2^{\ell}} ,
\label{eq:share}
\end{equation}
server $P_0$ holds $x_0$, and server $P_1$ holds $x_1$. We write $\share{x} = (x_0, x_1)$.

Since $x_0$ is uniformly random, $x_1 = x - x_0$ is also uniformly random, no matter what $x$ is. So each share, on its own, has the same distribution for every possible secret, and reveals nothing about $x$. Only the two shares together reveal $x$. To \emph{open} a shared value, each server sends its share to the other, which costs one round. We call this protocol $\pi_{\textsc{Open}}$.

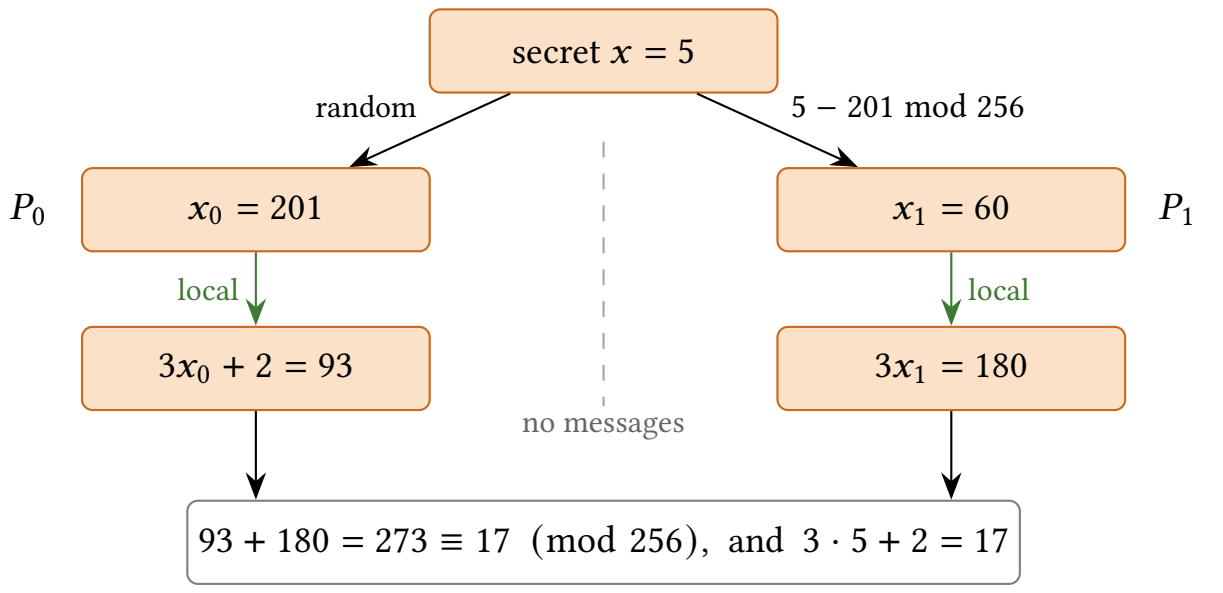
\begin{figure}[t]
\centering
\begin{adjustbox}{max width=\columnwidth}
\begin{tikzpicture}[font=\footnotesize, >={Stealth[length=1.8mm]},
  v/.style={draw, rounded corners=2pt, minimum width=2.3cm, minimum height=0.55cm, align=center, inner sep=2pt},
  sec/.style={v, fill=secfill, draw=secline},
  loc/.style={v, fill=locfill, draw=locline},
  lab/.style={font=\scriptsize}]
\node[sec] (x)  at (0,0)       {secret $x = 5$};
\node[sec] (x0) at (-2.3,-1.05) {$x_0 = 201$};
\node[sec] (x1) at (2.3,-1.05)  {$x_1 = 60$};
\node[font=\footnotesize\bfseries, left=3pt of x0] {$P_0$};
\node[font=\footnotesize\bfseries, right=3pt of x1] {$P_1$};
\draw[->] (x) -- node[above left=-1pt, lab] {random} (x0);
\draw[->] (x) -- node[above right=-1pt, lab] {$5 - 201 \bmod 256$} (x1);
\node[sec] (y0) at (-2.3,-2.1) {$3 x_0 + 2 = 93$};
\node[sec] (y1) at (2.3,-2.1)  {$3 x_1 = 180$};
\draw[->, locline] (x0) -- node[left, lab] {local} (y0);
\draw[->, locline] (x1) -- node[right, lab] {local} (y1);
\draw[dashed, black!35] (0,-0.6) -- (0,-2.35) node[below=-1pt, lab, text=black!60] {no messages};
\node[v, draw=black!50] (y) at (0,-3.25) {$93 + 180 = 273 \equiv 17 \pmod{256}$, \ and \ $3 \cdot 5 + 2 = 17$};
\draw[->] (y0) -- (y0 |- y.north);
\draw[->] (y1) -- (y1 |- y.north);
\end{tikzpicture}
\end{adjustbox}
\caption{Additive secret sharing in $\mathbb{Z}_{256}$. The secret $x = 5$ is
split into two shares that look random on their own. To compute the linear
function $y = 3x + 2$, each server applies it to its own share (only $P_0$ adds
the constant). The two results add up to the correct answer, without any
communication.}
\label{fig:ass}
\end{figure}

Figure~\ref{fig:ass} shows a small example with $\ell = 8$, so that all values live in $\mathbb{Z}_{256}$. To share $x = 5$, we draw a random $x_0 = 201$ and set $x_1 = 5 - 201 \bmod 256 = 60$. Indeed, $201 + 60 = 261 \equiv 5 \pmod{256}$. To compute $y = 3x + 2$, each server works on its own share. $P_0$ computes $3 \cdot 201 + 2 \bmod 256 = 93$, and $P_1$ computes $3 \cdot 60 \bmod 256 = 180$. The results $93$ and $180$ are a valid sharing of $3 \cdot 5 + 2 = 17$, since $93 + 180 = 273 \equiv 17 \pmod{256}$. No message was exchanged.

\subsection{Fixed-Point Encoding}

Neural networks compute with real numbers, while our protocols compute with integers modulo $2^{\ell}$. Secure computation on floating-point numbers is possible, but expensive. Instead, we map each real number to an integer with the commonly used \emph{fixed-point encoding}~\citep{catrina2010secure}, and then secret-share the integer. With $f$ fractional bits, a real number $a$ is represented by the integer
\begin{equation}
\bar a = \lfloor a \cdot 2^{f} \rceil \bmod 2^{\ell} ,
\label{eq:fixed}
\end{equation}
where $\lfloor \cdot \rceil$ denotes rounding to the nearest integer. We use $f = 16$. For example, $a = 1.5$ is encoded as $1.5 \cdot 2^{16} = 98{,}304$. Negative numbers use the two's complement convention. The integers from $0$ to $2^{\ell-1} - 1$ represent non-negative values, and the integers from $2^{\ell-1}$ to $2^{\ell} - 1$ represent negative values. For example, in $\mathbb{Z}_{256}$, the value $-5$ is stored as $256 - 5 = 251$. A useful consequence is that the sign of a value is its most significant bit, that is, $\msb(x) = 1$ if and only if $x$ is negative.

When two fixed-point values are multiplied, the product carries $2f$ fractional bits instead of $f$. To return to $f$ bits, the product must be divided by $2^{f}$, an operation called \emph{truncation}. On shares, truncation can be done locally~\citep{mohassel2017secureml}, with a protocol that we call $\pi_{\textsc{Trunc}}$. $P_0$ shifts its share to the right by $f$ bits, and $P_1$ does the same to the negation of its share and negates the result. The outcome is correct up to the least significant bit, except with a small probability that depends on how far the value is from the edges of the ring. Appendix~\ref{app:trunc} gives the details.

\subsection{Local and Interactive Operations}

Additive secret sharing is linear. This single property decides the cost of every layer of a neural network. Three operations cost no communication at all, because each server can carry them out on its own share,
\begin{align}
\share{x} + \share{y} &= (x_0 + y_0,\; x_1 + y_1), \label{eq:add}\\
c + \share{x} &= (c + x_0,\; x_1), \label{eq:addc}\\
c \cdot \share{x} &= (c\, x_0,\; c\, x_1), \label{eq:mulc}
\end{align}
where $c$ is a public constant. It follows that any linear function with public coefficients is free. In particular, if $W$ is a public weight matrix and $X$ is a secret matrix, then
\begin{equation}
W \share{X} = (W X_0,\; W X_1) ,
\label{eq:matmul}
\end{equation}
so a matrix multiplication or a convolution with public weights, as in our setting, is computed locally by each server. We call this protocol $\pi_{\textsc{MatMul}}$, and describe how we run it on a GPU in Section~\ref{sec:secure}.

\paragraph{Multiplication of shared values.} Multiplication of two shared values, however, cannot be done locally, and requires the servers to interact. Expanding the product gives
\begin{equation}
x \cdot y = (x_0 + x_1)(y_0 + y_1) = x_0 y_0 + x_0 y_1 + x_1 y_0 + x_1 y_1 .
\label{eq:expand}
\end{equation}
$P_0$ can compute $x_0 y_0$ and $P_1$ can compute $x_1 y_1$, but no single server holds both factors of the cross terms $x_0 y_1$ and $x_1 y_0$. The standard solution uses a \emph{Beaver triple}~\citep{beaver1991efficient}, a sharing of random values $a$, $b$, and $c = a \cdot b$ prepared by the dealer. The servers open $d = x - a$ and $e = y - b$, which reveal nothing because $a$ and $b$ are random, and then compute locally
\begin{equation}
\share{x \cdot y} = \share{c} + d \cdot \share{b} + e \cdot \share{a} + d \cdot e .
\label{eq:beaver}
\end{equation}
A secret-by-secret multiplication, which we call $\pi_{\textsc{Mul}}$, therefore costs one round. Since the weights are public in our setting, the multiplications in dot products, matrix products, and convolutions are all multiplications by public values, and each server computes them locally (Equation~\eqref{eq:matmul}). Secret-by-secret multiplications appear only inside the non-linear layers. A ReLU, for example, multiplies its secret input by a secret bit that tells whether the input is non-negative (Section~\ref{sec:fss}).

Non-linear functions are the expensive part of secure inference. A ReLU must decide whether its input is negative, and a max-pooling layer must decide which of two values is larger. Both are comparisons, which cannot be done locally, as we explain in Section~\ref{sec:compare}.

\subsection{Secure Comparison}
\label{sec:compare}

In the two's complement representation of $x$, the sign of $x$ is given by its most significant bit. A server cannot read the sign from its own share. Consider $\mathbb{Z}_{256}$, and suppose $P_0$ holds $x_0 = 201$. If the secret is $x = 5$, then $P_1$ holds $x_1 = 60$. If the secret is $x = -5$, which is stored as $251$, then $P_1$ holds $x_1 = 50$. In both cases $P_0$ sees exactly the same share, yet the signs are opposite. The sign of $x = x_0 + x_1$ depends on the carry that ripples up from the lower bits of the addition, and computing this carry requires information from both servers.

The classic approach computes the carry explicitly. The servers convert their shares into shares of individual bits, and then evaluate a Boolean adder circuit on the bits. Each AND gate in the circuit needs communication, and a ripple-carry adder processes the bits one after another, which takes about $\ell$ rounds. Faster adders reduce the rounds, but send more bytes. Preprocessed random bits, such as edaBits~\citep{escudero2020improved}, make the conversion cheaper, but the cost remains high in both rounds and bytes. We therefore use function secret sharing, described next.

\subsection{Function Secret Sharing}
\label{sec:fss}

Additive secret sharing hides a \emph{value}, while function secret sharing (FSS)~\citep{boyle2015function, boyle2016function} hides a \emph{function}. In the offline phase, the dealer splits a function $g$ into two \emph{keys}, $k_0$ and $k_1$, and gives $k_0$ to $P_0$ and $k_1$ to $P_1$. FSS comes with a public evaluation algorithm, $\mathrm{Eval}$, that takes a key and an input and returns a number. The keys are built so that, for every input $z$,
\begin{equation}
\mathrm{Eval}(k_0, z) + \mathrm{Eval}(k_1, z) = g(z) .
\label{eq:fss}
\end{equation}
Each server runs $\mathrm{Eval}$ on its own key, locally, and the two outputs form an additive sharing of $g(z)$. Moreover, each key on its own looks random and reveals nothing about $g$. There is one catch. Both servers must evaluate their keys on the same input $z$, so $z$ must be public. FSS hides the function, not the input.

The function we need is a \emph{comparison}. A distributed comparison function (DCF)~\citep{boyle2016function, boyle2021function} is an FSS scheme for the comparison with a secret threshold $\alpha$,
\begin{equation}
g_{\alpha, \beta}(z) =
\begin{cases}
\beta & \text{if } z < \alpha, \\
0 & \text{otherwise,}
\end{cases}
\label{eq:dcf}
\end{equation}
where the output $\beta$ is also secret. To see how the keys can hide $\alpha$ and $\beta$, consider the simplest possible scheme, which writes the function as a table. Take 3-bit inputs $z \in \{0, \dots, 7\}$, the threshold $\alpha = 6$, and $\beta = 1$, so that $g$ is the table of its eight outputs. The dealer secret-shares this table entry by entry in $\mathbb{Z}_{256}$, exactly as in Section~\ref{sec:ass}. It draws a random table $k_0$ and sets $k_1 = g - k_0 \bmod 256$:
\begin{equation}
\begin{array}{r|cccccccc}
z & 0 & 1 & 2 & 3 & 4 & 5 & 6 & 7 \\
\hline
g(z) & 1 & 1 & 1 & 1 & 1 & 1 & 0 & 0 \\
k_0 & 17 & 200 & 31 & 94 & 250 & 3 & 77 & 140 \\
k_1 & 240 & 57 & 226 & 163 & 7 & 254 & 179 & 116
\end{array}
\label{eq:table}
\end{equation}
Here $\mathrm{Eval}(k_i, z)$ simply reads entry $z$ of table $k_i$. For the public input $z = 5$, $P_0$ reads $3$ and $P_1$ reads $254$, and $3 + 254 = 257 \equiv 1 = g(5) \pmod{256}$. Each table on its own is uniformly random, so a server cannot tell where the outputs switch from $1$ to $0$, and learns neither $\alpha$ nor $\beta$. The threshold is hidden in the keys in the same way that a secret is hidden in its shares. The problem with this scheme is size. A table has one entry for every possible input, and our comparisons are on 63-bit inputs, which would need $2^{63}$ entries. A DCF compresses each table into a short key, of roughly one 128-bit word per input bit, while keeping both properties.

\paragraph{Computing the sign with a DCF.} We use a DCF to compute the sign of a shared value $\share{x}$. The difficulty is that FSS needs a public input, while $x$ is secret. We solve this with the same idea as the Beaver triple, a random mask. In the offline phase, the dealer draws a random mask $r \in \ring$ and gives the servers a sharing $\share{r}$. In the online phase, the servers open the masked value $\hat x = x + r$. Since $r$ is uniformly random, $\hat x$ is uniformly random too, and reveals nothing about $x$. Now $\hat x$ is public, and $x = \hat x - r$. If the servers knew $r$, they could simply subtract. They do not, because only the dealer knows $r$. However, the sign of $\hat x - r$ can be written as
\begin{equation}
\msb(x) \;=\; \msb(\hat x) \;\oplus\; \msb(r) \;\oplus\; \ind{\xlow < \rlow} ,
\label{eq:msb}
\end{equation}
where $\oplus$ is exclusive OR (XOR), and $\xlow$ and $\rlow$ are the lower $\ell - 1$ bits of $\hat x$ and $r$, that is, all bits except the most significant one. The identity is ordinary subtraction in binary. Subtracting $r$ from $\hat x$ flips the top bit of $\hat x$ if the top bit of $r$ is $1$, and flips it once more if the subtraction of the lower bits borrows, which happens exactly when $\xlow < \rlow$. Each server can compute every term of Equation~\eqref{eq:msb}. The first term is public. For the second term, the dealer gives the servers XOR shares of $\msb(r)$. The third term compares the public value $\xlow$ with the secret threshold $\rlow$, which is exactly a DCF with $\alpha = \rlow$ and $\beta = 1$, evaluated at a public input. The dealer gives $P_0$ and $P_1$ the two keys of this DCF, and each server evaluates its key on $\xlow$ locally, with no communication. The outputs of this DCF are bits, so the servers obtain XOR shares of $\ind{\xlow < \rlow}$, where XOR is addition modulo~$2$. Combining the three terms with XOR, each server obtains an XOR share of $\msb(x)$.

For example, take $\ell = 8$, so the lower bits are the lower 7 bits. Let the secret be $x = -5$, stored as $251$, and let the mask be $r = 100$. The servers open $\hat x = 251 + 100 \bmod 256 = 95$. Then $\msb(\hat x) = 0$, $\msb(r) = 0$, and $\xlow = 95 < 100 = \rlow$, so the last term is $1$. Equation~\eqref{eq:msb} gives $\msb(x) = 0 \oplus 0 \oplus 1 = 1$, so $x$ is negative, as expected. Had the secret been $x = 5$, the servers would have opened $\hat x = 105$, the last term would have been $0$, and the result would have been $\msb(x) = 0$. In both cases the servers saw only a random-looking $\hat x$.

\begin{figure}[!htb]
\centering
\begin{adjustbox}{max width=\columnwidth}
\begin{tikzpicture}[font=\small, >={Stealth[length=2mm]}, node distance=3.5mm,
  st/.style={draw, rounded corners=2pt, text width=6.1cm, font=\footnotesize, align=left, inner sep=5pt},
  off/.style={st, fill=black!3, draw=black!50, dashed},
  com/.style={st, fill=comfill, draw=comline},
  loc/.style={st, fill=locfill, draw=locline},
  tag/.style={font=\scriptsize\bfseries, anchor=west, xshift=4pt}]
\node[off] (s0) {\textbf{Offline (dealer).} Draw a random mask $r$. Give server
$P_i$ a share $r_i$ of $r$, a share of $\msb(r)$, a DCF key $k_i$ for the
threshold $\alpha = \rlow$, and a Beaver triple.};
\node[com, below=of s0] (s1) {\textbf{1. Mask and open.} Each $P_i$ sends
$x_i + r_i$. Both learn $\hat x = x + r$, which is uniformly random.};
\node[loc, below=of s1] (s2) {\textbf{2. Compare.} Each $P_i$ evaluates its key
on the public $\xlow$ and obtains a share of $\ind{\xlow < \rlow}$.};
\node[loc, below=of s2] (s3) {\textbf{3. Combine.} Using Equation~\eqref{eq:msb},
each $P_i$ obtains an XOR share of $\msb(x)$, and hence of
$b = \ind{x \ge 0} = 1 \oplus \msb(x)$.};
\node[com, below=of s3] (s4) {\textbf{4. Convert and multiply.} Convert $b$ to
additive shares with Equation~\eqref{eq:b2a}, then compute
$\relu(x) = b \cdot x$.};
\foreach \a/\b in {s0/s1, s1/s2, s2/s3, s3/s4} \draw[->, black!50] (\a) -- (\b);
\node[tag, black!55] at (s0.east) {untimed};
\node[tag, comline] at (s1.east) {1 round};
\node[tag, locline] at (s2.east) {local};
\node[tag, locline] at (s3.east) {local};
\node[tag, comline] at (s4.east) {2 rounds};
\end{tikzpicture}
\end{adjustbox}
\caption{A secure ReLU with function secret sharing. Red steps communicate and
green steps are local. The expensive part of a comparison, step 2, is a local
computation on the GPU. The servers only ever see the random-looking value
$\hat x$.}
\label{fig:fss}
\end{figure}
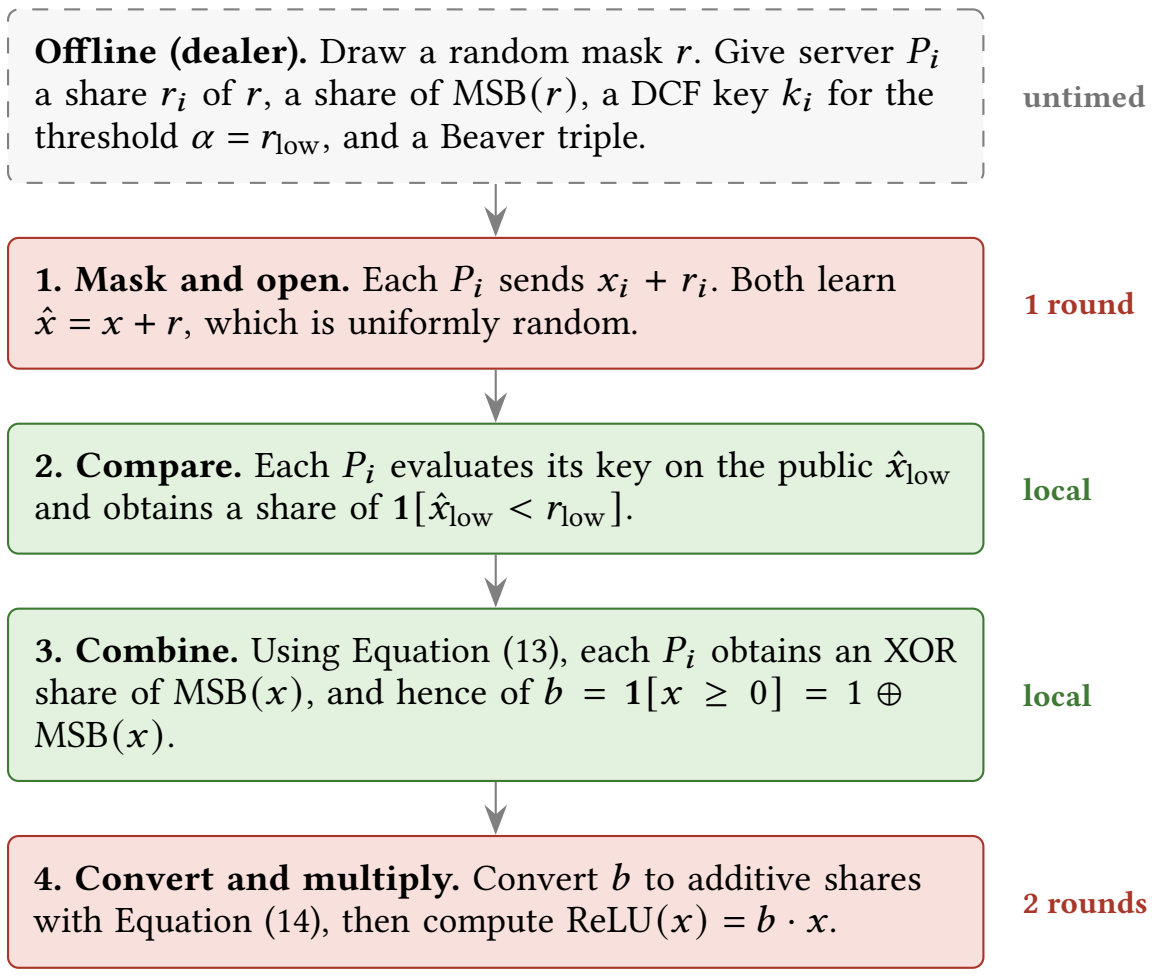

A ReLU needs the bit $b = \ind{x \ge 0} = 1 \oplus \msb(x)$, which tells whether $x$ is non-negative. The steps above, which we call $\pi_{\textsc{Sign}}$, give the servers XOR shares of $b$, that is, bits $b_0$ and $b_1$ with $b = b_0 \oplus b_1$. To use $b$ in arithmetic, we convert it into additive shares with a protocol that we call $\pi_{\textsc{B2A}}$,
\begin{equation}
b \;=\; b_0 \oplus b_1 \;=\; b_0 + b_1 - 2\, b_0 b_1 ,
\label{eq:b2a}
\end{equation}
where $b_0$ is known to $P_0$ and $b_1$ to $P_1$, so the product $b_0 b_1$ is one Beaver multiplication. Then
\begin{equation}
\relu(x) = b \cdot x , \qquad \max(u, v) = u + \relu(v - u) .
\label{eq:relu}
\end{equation}
Algorithm~\ref{alg:relu} puts these steps together into the protocol $\pi_{\textsc{ReLU}}$, and Figure~\ref{fig:fss} illustrates it. In the offline phase, the dealer gives each server its share of the mask $r$, its XOR share of $\msb(r)$, its DCF key, and the Beaver triples for the two multiplications. Online, line~1 opens the masked value $\hat x$, which costs one round. Line~2 evaluates the DCF key on the public $\xlow$, locally. Line~3 combines the three terms of Equation~\eqref{eq:msb} into an XOR share of $b = 1 \oplus \msb(x)$. Each server XORs its DCF output $d_i$ and its share $m_i$ of $\msb(r)$, and only $P_0$ also XORs in the public bit $\msb(\hat x) \oplus 1$, so that this bit is counted once. Lines~1 to~3 form $\pi_{\textsc{Sign}}$. Line~4 converts $b$ into additive shares with Equation~\eqref{eq:b2a}, and line~5 multiplies it with $x$. Each of these two lines is one Beaver multiplication and costs one round, so a ReLU costs three rounds in total.

\begin{algorithm}[t]
\caption{$\pi_{\textsc{ReLU}}(\share{x})$, secure ReLU with FSS.}
\label{alg:relu}
\begin{algorithmic}[1]
\Statex \textbf{Offline (dealer):} draw $r \in \ring$; give $P_i$ its share $r_i$, an XOR share $m_i$ of $\msb(r)$, the DCF key $k_i$ for $\alpha = \rlow$ and $\beta = 1$, and two Beaver triples.
\Statex \textbf{Online (server $P_i$, $i \in \{0, 1\}$):}
\State $\hat x \gets \pi_{\textsc{Open}}(\share{x} + \share{r})$ \Comment{round 1}
\State $d_i \gets \mathrm{Eval}(k_i, \xlow)$ \Comment{local}
\State $b_i \gets d_i \oplus m_i \oplus (1 - i) \cdot \big(\msb(\hat x) \oplus 1\big)$ \Comment{local}
\State $\share{b} \gets \pi_{\textsc{B2A}}(b_0, b_1)$ \Comment{round 2, Eq.~\eqref{eq:b2a}}
\State \Return $\pi_{\textsc{Mul}}(\share{b}, \share{x})$ \Comment{round 3}
\end{algorithmic}
\end{algorithm}

Max-pooling reuses the ReLU. The maximum of two values, which we call $\pi_{\textsc{Max}}$, is the first value plus the ReLU of their difference (Equation~\eqref{eq:relu}), so it costs the same as a ReLU. To take the maximum of $L$ values, which we call $\pi_{\textsc{MaxPool}}$, we compare them in pairs, in a tree. For example, with four values, the first level computes $\max(x_1, x_2)$ and $\max(x_3, x_4)$ at the same time, and the second level takes the maximum of the two results. In general, the tree has $\lceil \log_2 L \rceil$ levels and $L - 1$ comparisons in total. All comparisons in the same level are evaluated together, so they share the same three rounds, and so do all samples in a batch.

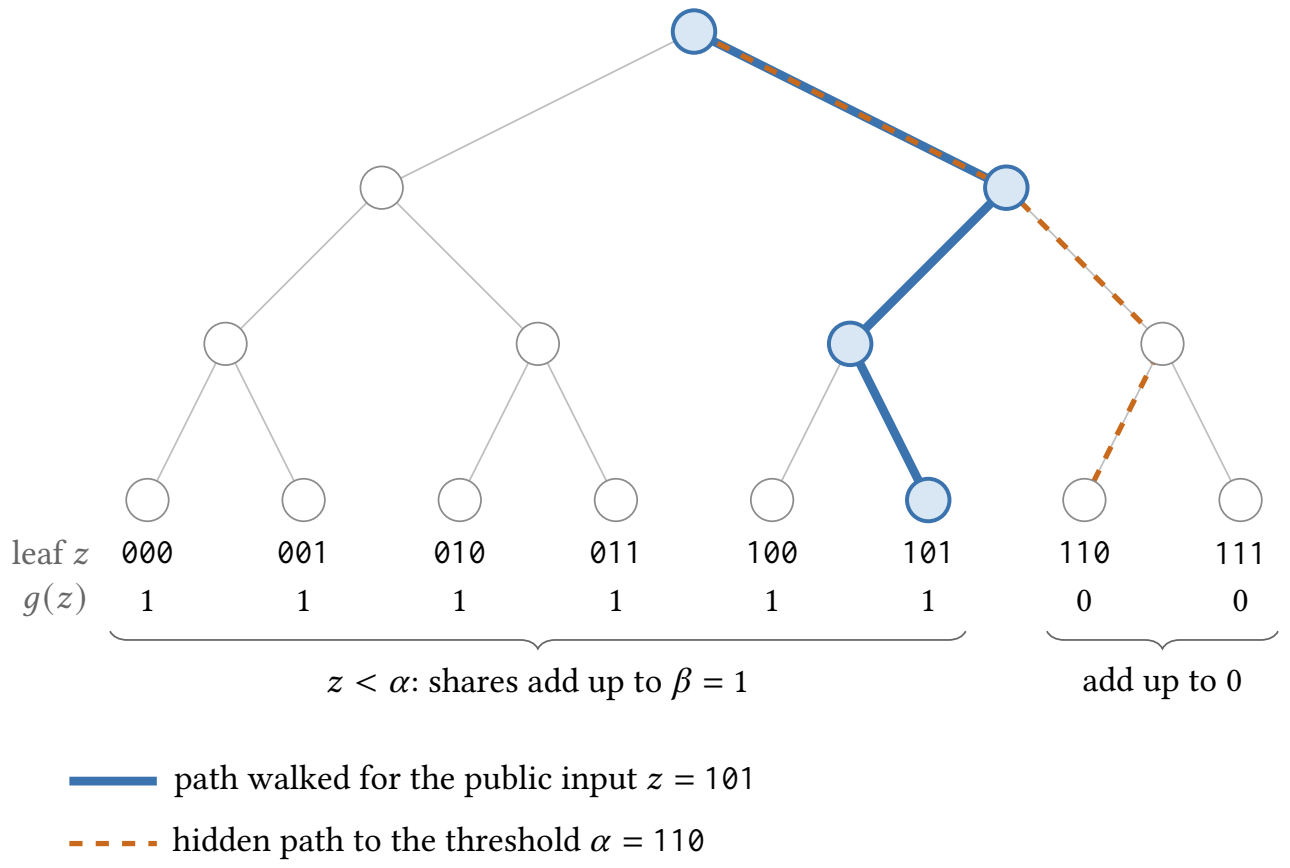
\begin{figure}[!htb]
\centering
\begin{adjustbox}{max width=\columnwidth}
\begin{tikzpicture}[font=\small]
  \foreach \l in {0,1,2,3}{
    \pgfmathtruncatemacro{\n}{2^\l-1}
    \foreach \i in {0,...,\n}{
      \pgfmathsetmacro{\xx}{(\i+0.5)*10/(2^\l)-5}
      \coordinate (n\l-\i) at (\xx,{-1.25*\l});
    }
  }
  \foreach \l in {1,2,3}{
    \pgfmathtruncatemacro{\n}{2^\l-1}
    \pgfmathtruncatemacro{\pl}{\l-1}
    \foreach \i in {0,...,\n}{
      \pgfmathtruncatemacro{\p}{floor(\i/2)}
      \draw[black!25] (n\pl-\p) -- (n\l-\i);
    }
  }
  \draw[publine, line width=2pt] (n0-0) -- (n1-1) -- (n2-2) -- (n3-5);
  \draw[secline, line width=1.2pt, dashed] (n0-0) -- (n1-1) -- (n2-3) -- (n3-6);
  \foreach \l in {0,1,2,3}{
    \pgfmathtruncatemacro{\n}{2^\l-1}
    \foreach \i in {0,...,\n}{
      \node[circle, draw=black!45, fill=white, minimum size=3.4mm, inner sep=0pt] at (n\l-\i) {};
    }
  }
  \foreach \c in {n0-0,n1-1,n2-2,n3-5}
    \node[circle, draw=publine, fill=pubfill, line width=0.9pt, minimum size=3.4mm, inner sep=0pt] at (\c) {};
  \foreach \i/\b/\v in {0/000/1,1/001/1,2/010/1,3/011/1,4/100/1,5/101/1,6/110/0,7/111/0}{
    \node at ($(n3-\i)+(0,-0.42)$) {\texttt{\b}};
    \node at ($(n3-\i)+(0,-0.80)$) {$\v$};
  }
  \node[anchor=east, text=black!60] at ($(n3-0)+(-0.35,-0.42)$) {leaf $z$};
  \node[anchor=east, text=black!60] at ($(n3-0)+(-0.35,-0.80)$) {$g(z)$};
  \draw[decorate, decoration={brace, mirror, amplitude=4pt}, black!60]
    ($(n3-0)+(-0.3,-1.05)$) -- ($(n3-5)+(0.3,-1.05)$)
    node[midway, below=5pt, text=black] {$z < \alpha$: shares add up to $\beta = 1$};
  \draw[decorate, decoration={brace, mirror, amplitude=4pt}, black!60]
    ($(n3-6)+(-0.3,-1.05)$) -- ($(n3-7)+(0.3,-1.05)$)
    node[midway, below=5pt, text=black] {add up to $0$};
  \draw[publine, line width=2pt] (-5,-6.0) -- (-4.3,-6.0)
    node[right, black] {path walked for the public input $z = \texttt{101}$};
  \draw[secline, line width=1.2pt, dashed] (-5,-6.5) -- (-4.3,-6.5)
    node[right, black] {hidden path to the threshold $\alpha = \texttt{110}$};
\end{tikzpicture}
\end{adjustbox}
\caption{Evaluating a distributed comparison function on 3-bit inputs. The keys
define a binary tree whose path to the secret threshold $\alpha$ (orange, dashed)
separates the leaves with $z < \alpha$ from the rest. A server evaluates its key
at a public input $z$ by walking a single root-to-leaf path (blue). The two
servers' outputs add up to $g(z) = \ind{z < \alpha}$, here $1$, yet neither
server learns $\alpha$. In our solution the tree has 63 levels.}
\label{fig:dcf}
\end{figure}

\paragraph{DCF construction.} A DCF replaces the table, which would have $2^{63}$ entries, with a short key. The construction~\citep{boyle2016function, boyle2021function} is built on a binary tree of pseudorandom values, in the style of \citet{goldreich1986construct}. The leaves of the tree are the $2^{n}$ possible inputs, where $n = \ell - 1 = 63$ in our case, and the path from the root to the leaf $\alpha$ splits the leaves into those smaller than $\alpha$ and the rest (Figure~\ref{fig:dcf}). Each server's key contains a random 128-bit root seed and one short \emph{correction word} per level of the tree. A pseudorandom generator (PRG) expands a seed into the seeds of its two children. The correction words are chosen so that the two servers' trees are identical everywhere except on the path to $\alpha$, and so that small value words along the path add up to $\beta$ exactly for the leaves left of the path.

To evaluate its key at an input $z$, a server does not visit all $2^n$ leaves. It walks a single path from the root to the leaf $z$, turning left or right at each level according to the next bit of $z$. At each level, it expands its current seed with the PRG into two child seeds, keeps the child on its path, and corrects it with the correction word of that level. Along the way, it adds up a small output word. As long as the path to $z$ follows the path to $\alpha$, the two servers hold different seeds, and the comparison is not yet decided. At the first level where the two paths split, the comparison is decided. If $z$ turns left where $\alpha$ turns right, then $z < \alpha$, and the correction words make the two servers' output words at this level add up to $\beta$. In the other case, $z > \alpha$, and they add up to $0$. From that level on, the correction words make the two servers' seeds identical, so all their later output words are equal and cancel when the shares are combined. Since each server starts from its own random root seed, its walk looks random on its own, and its key reveals nothing about $\alpha$. An evaluation costs one PRG expansion per level, so a comparison on 63 bits takes 63 steps of local computation, one after another. Appendix~\ref{app:dcf} gives the full key generation and evaluation algorithms.

Our PRG is built from AES-128 in a fixed-key construction. For a seed $s$,
\begin{equation}
H(s) = \mathrm{AES}_K(s) \oplus s ,
\label{eq:mmo}
\end{equation}
where $K$ is a fixed public key. This is the Matyas--Meyer--Oseas construction. Fixed-key AES is a standard and efficient way to build such functions in MPC~\citep{guo2020efficient}. The seeds are 128 bits long, which meets the 128-bit security requirement. Each PRG expansion calls $H$ twice, once to derive the child seed and once to derive the value word.

FSS thus moves the heavy work of a comparison into the offline key generation, which is not timed, and into the local key evaluation, which needs no communication and suits a GPU. Online, a ReLU sends only one masked value and two Beaver multiplications, far less than bit decomposition in both rounds and bytes. The price is computation, 63 levels of AES for each of thousands of comparisons per sample. Making this computation fast is one of the GPU problems we address in Section~\ref{sec:secure}.

Table~\ref{tab:cost} lists the protocols that we use, with their online cost, in three levels. The basic primitives operate on shared values. The building blocks are built from the primitives, and the network layers are built from the building blocks. For example, $\pi_{\textsc{Conv}}$ uses $\pi_{\textsc{MatMul}}$, which uses local multiplications by public values and $\pi_{\textsc{Trunc}}$, and $\pi_{\textsc{ReLU}}$ uses $\pi_{\textsc{Sign}}$, $\pi_{\textsc{B2A}}$, and $\pi_{\textsc{Mul}}$. Protocols with zero rounds are local, and the others need communication.

\begin{table*}[t]
\centering
\caption{The protocols used in our solution and their online cost, in three levels. Bits are those sent by each server per output element, with $\ell = 64$. $W$, $b$, and $c$ are public, and $L$ is the number of values in a max-pooling window.}
\label{tab:cost}
\small
\begin{tabular}{@{}llccl@{}}
\toprule
Protocol & Description & Rounds & Bits sent & Reference \\
\midrule
\multicolumn{5}{@{}l}{\textit{Basic primitives}} \\
$\share{x} + \share{y}$, \ $c + \share{x}$, \ $c\share{x}$ & addition, and operations with a public constant & 0 & 0 & Eqs.~\eqref{eq:add}--\eqref{eq:mulc} \\
$\pi_{\textsc{Trunc}}$ & division by $2^{f}$ after a product & 0 & 0 & \citet{mohassel2017secureml} \\
$\pi_{\textsc{Open}}$ & reveal $x$ to both servers & 1 & $\ell$ & --- \\
$\pi_{\textsc{Mul}}$ & product $\share{x \cdot y}$ with a Beaver triple & 1 & $2\ell$ & \citet{beaver1991efficient} \\
$\pi_{\textsc{Sign}}$ & XOR shares of $\ind{x \ge 0}$, by FSS & 1 & $\ell$ & \citet{boyle2021function} \\
$\pi_{\textsc{B2A}}$ & XOR shares to additive shares, Eq.~\eqref{eq:b2a} & 1 & $2\ell$ & \citet{boyle2021function} \\
\midrule
\multicolumn{5}{@{}l}{\textit{Building blocks}} \\
$\pi_{\textsc{MatMul}}$ & matrix product $\share{X} W$ with public $W$, then $\pi_{\textsc{Trunc}}$ & 0 & 0 & Eq.~\eqref{eq:matmul}, Alg.~\ref{alg:matmul} \\
$\pi_{\textsc{ReLU}}$ & $\share{\max(x, 0)}$, from $\pi_{\textsc{Sign}}$, $\pi_{\textsc{B2A}}$, and $\pi_{\textsc{Mul}}$ & 3 & $5\ell$ & \citet{boyle2021function}, Alg.~\ref{alg:relu} \\
$\pi_{\textsc{Max}}$ & $\share{\max(u, v)} = \share{u} + \pi_{\textsc{ReLU}}(\share{v} - \share{u})$ & 3 & $5\ell$ & Eq.~\eqref{eq:relu} \\
\midrule
\multicolumn{5}{@{}l}{\textit{Network layers}} \\
$\pi_{\textsc{FC}}$ & fully connected layer $\share{X} W + b$, via $\pi_{\textsc{MatMul}}$ & 0 & 0 & Section~\ref{sec:secure} \\
$\pi_{\textsc{Conv}}$ & convolution with public filters, via $\pi_{\textsc{MatMul}}$ & 0 & 0 & Section~\ref{sec:secure} \\
$\pi_{\textsc{MaxPool}}$ & maximum of $L$ values, tree of $\pi_{\textsc{Max}}$ & $3\lceil \log_2 L \rceil$ & $5\ell(L-1)$ & Section~\ref{sec:fss} \\
\bottomrule
\end{tabular}
\end{table*}

\section{\model: An MPC-Friendly DTI Model}
\label{sec:model}

The competition provided DeepDTAGen~\citep{shah2025deepdtagen} as the reference model and allowed teams to change both the architecture and the weights, as long as the accuracy stayed within 2\% of the reference. The affinity branch of DeepDTAGen represents the drug as a molecular graph and processes it with graph convolutions. In a graph convolution, the features of each atom are combined with those of its neighbors, which means multiplying the feature matrix by the adjacency matrix of the graph. Both matrices depend on the drug, so both are secret, and every such product is a secret-by-secret multiplication that costs a round and sends data proportional to the size of the matrices. The protein is processed by a gated convolutional network whose output is flattened into a fully connected layer with 1.3 million parameters. In total, the affinity branch has 3.58 million parameters. Although the protein side is public in our setting, the drug side and the head still require many secret products and comparisons. To reduce this cost, we designed an MPC-friendly model, \model.

Section~\ref{sec:primer} showed that linear operations with public weights are local, secret-by-secret products cost a round, and comparisons cost several rounds and most of the bytes. The key ideas in designing \model were the following.
\begin{enumerate}[leftmargin=1.6em, itemsep=2pt]
\item Keep every secret operation linear in public weights. Then the operation is local (Equation~\eqref{eq:matmul}). This rules out graph convolutions and attention on the drug.
\item Count comparisons, not parameters. The number of ReLU and max comparisons on secret values, and the depth at which they occur, determine the online time. A model can have many parameters and still be cheap, if its secret activations are few.
\item Spend capacity where it is free. The protein tower is public, so making it wider or deeper costs no secure computation. It only adds plaintext GPU work, which is fast.
\item Use only functions that MPC computes exactly. ReLU and maximum are computed exactly by FSS. Functions such as sigmoid, softmax, or division need approximations, which add both cost and error.
\end{enumerate}

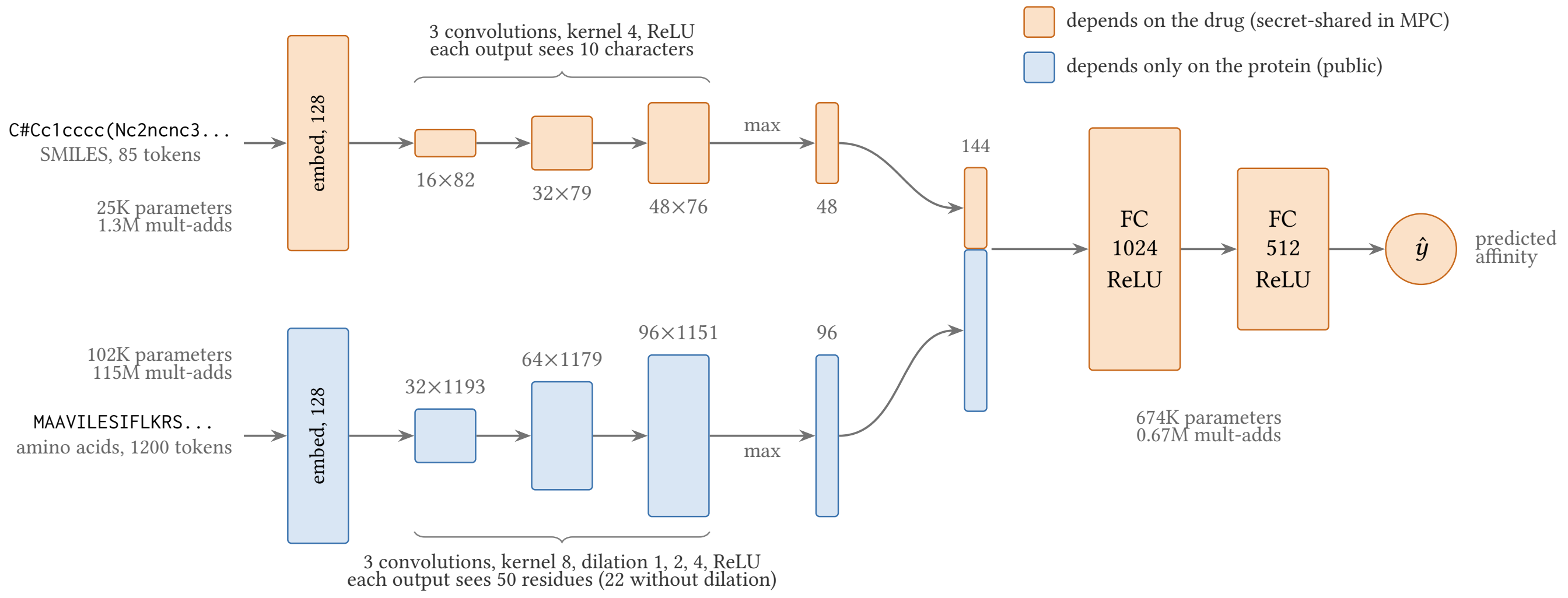
\begin{figure*}[t]
\centering
\resizebox{\textwidth}{!}{%
\begin{tikzpicture}[
  x=1.05cm, y=1cm, font=\footnotesize, >={Stealth[length=1.6mm]},
  blk/.style={draw, rounded corners=1pt, minimum width=6mm, inner sep=0pt},
  sec/.style={blk, fill=secfill, draw=secline},
  pub/.style={blk, fill=pubfill, draw=publine},
  dimtxt/.style={font=\scriptsize, text=black!60},
  arr/.style={->, semithick, black!55},
]
\def\yd{1.45}\def\yp{-1.45}
\node[align=center, anchor=east] (din) at (2.1,\yd) {\ttfamily\scriptsize C\#Cc1cccc(Nc2ncnc3...\\[-1pt]{\rmfamily\scriptsize\color{black!60}SMILES, 85 tokens}};
\node[sec, minimum height=2.13cm] (de) at (2.8,\yd) {\rotatebox{90}{\scriptsize embed, 128}};
\node[sec, minimum height=0.27cm] (d1) at (4.0,\yd) {};
\node[sec, minimum height=0.53cm] (d2) at (5.1,\yd) {};
\node[sec, minimum height=0.80cm] (d3) at (6.2,\yd) {};
\node[sec, minimum width=2.2mm, minimum height=0.80cm] (dv) at (7.6,\yd) {};
\foreach \a/\b in {din/de, de/d1, d1/d2, d2/d3} \draw[arr] (\a) -- (\b);
\draw[arr] (d3) -- node[above, dimtxt]{max} (dv);
\foreach \n/\t in {d1/16{\times}82, d2/32{\times}79, d3/48{\times}76, dv/48}
  \node[dimtxt, below=1pt of \n] {$\t$};
\draw[decorate, decoration={brace, amplitude=3pt}, black!50] ($(d1.west)+(0,0.6)$) -- ($(d3.east)+(0,0.6)$)
  node[midway, above=3pt, dimtxt, text=black!80, align=center] {3 convolutions, kernel 4, ReLU\\[-2pt]each output sees 10 characters};
\node[align=center, anchor=east] (pin) at (2.1,\yp) {\ttfamily\scriptsize MAAVILESIFLKRS...\\[-1pt]{\rmfamily\scriptsize\color{black!60}amino acids, 1200 tokens}};
\node[pub, minimum height=2.13cm] (pe) at (2.8,\yp) {\rotatebox{90}{\scriptsize embed, 128}};
\node[pub, minimum height=0.53cm] (p1) at (4.0,\yp) {};
\node[pub, minimum height=1.07cm] (p2) at (5.1,\yp) {};
\node[pub, minimum height=1.60cm] (p3) at (6.2,\yp) {};
\node[pub, minimum width=2.2mm, minimum height=1.60cm] (pv) at (7.6,\yp) {};
\foreach \a/\b in {pin/pe, pe/p1, p1/p2, p2/p3} \draw[arr] (\a) -- (\b);
\draw[arr] (p3) -- node[below, dimtxt]{max} (pv);
\foreach \n/\t in {p1/32{\times}1193, p2/64{\times}1179, p3/96{\times}1151, pv/96}
  \node[dimtxt, above=1pt of \n] {$\t$};
\draw[decorate, decoration={brace, mirror, amplitude=3pt}, black!50] ($(p1.west)+(0,-0.95)$) -- ($(p3.east)+(0,-0.95)$)
  node[midway, below=3pt, dimtxt, text=black!80, align=center] {3 convolutions, kernel 8, dilation 1, 2, 4, ReLU\\[-2pt]each output sees 50 residues (22 without dilation)};
\node[sec, minimum width=2.2mm, minimum height=0.80cm, anchor=south] (cd) at (9.0,0.4) {};
\node[pub, minimum width=2.2mm, minimum height=1.60cm, anchor=north] (cp) at (9.0,0.4) {};
\node[dimtxt, above=1pt of cd] {144};
\draw[arr] (dv.east) to[out=0, in=180] (cd.west);
\draw[arr] (pv.east) to[out=0, in=180] (cp.west);
\node[sec, minimum width=9mm, minimum height=2.4cm, align=center] (f1) at (10.5,0.4) {FC\\1024\\[1pt]ReLU};
\node[sec, minimum width=9mm, minimum height=1.6cm, align=center] (f2) at (11.9,0.4) {FC\\512\\[1pt]ReLU};
\node[sec, circle, minimum size=7mm, inner sep=0pt] (out) at (13.2,0.4) {$\hat y$};
\draw[arr] (9.15,0.4) -- (f1.west);
\draw[arr] (f1) -- (f2);
\draw[arr] (f2) -- (out);
\node[dimtxt, right=2pt of out, align=left] {predicted\\[-2pt]affinity};
\node[dimtxt, align=center] at (11.2,-1.35) {674K parameters\\[-2pt]0.67M mult-adds};
\node[dimtxt, anchor=north east, align=right] at (2.1,\yd-0.45) {25K parameters\\[-2pt]1.3M mult-adds};
\node[dimtxt, anchor=south east, align=right] at (2.1,\yp+0.45) {102K parameters\\[-2pt]115M mult-adds};
\node[sec, minimum width=3mm, minimum height=3mm] (l1) at (9.6,2.65) {};
\node[anchor=west, dimtxt, text=black!80] at (l1.east) {depends on the drug (secret-shared in MPC)};
\node[pub, minimum width=3mm, minimum height=3mm] (l2) at (9.6,2.2) {};
\node[anchor=west, dimtxt, text=black!80] at (l2.east) {depends only on the protein (public)};
\end{tikzpicture}%
}
\caption{\model, with shapes (channels${\times}$positions) for Davis. Block
heights are proportional to channel counts (fully connected layers are not to
scale), and ``max'' is a global maximum over positions. Orange marks everything
that depends on the drug, the only secret. Blue marks the protein tower, which
depends only on public information. The protein tower performs 98\% of the
plaintext multiply-adds, but needs no secure computation.}
\label{fig:model}
\end{figure*}

\model, shown in Figure~\ref{fig:model}, is a two-tower convolutional model in the style of DeepDTA~\citep{ozturk2018deepdta}. One tower reads the drug, the other reads the protein, and a small head combines the two into a predicted affinity. Table~\ref{tab:compare} compares \model with DeepDTA and DeepDTAGen. \model has 800,417 parameters, 4.5 times fewer than the affinity branch of DeepDTAGen, and it is as accurate on both Davis and KIBA (Section~\ref{sec:results}). We now describe each component and the reason behind it.

\begin{table*}[t]
\centering
\caption{\model compared with DeepDTA and DeepDTAGen. Balanced accuracy is the
mean over 7 affinity thresholds on the 5,010 Davis and 19,653 KIBA test pairs.
$^*$Measured by running the released DeepDTAGen checkpoint on the same split. The \model numbers are for the checkpoint we submitted, our best training run. Over training seeds, its Davis balanced accuracy averages about 0.82.}
\label{tab:compare}
\small
\begin{tabular}{@{}lccc@{}}
\toprule
 & DeepDTA~\citep{ozturk2018deepdta} & DeepDTAGen (affinity) & \model \\
\midrule
Drug input & SMILES & molecular graph & SMILES \\
Drug encoder & 3 conv (32, 64, 96) & 3 graph conv & 3 conv (16, 32, 48) \\
Protein encoder & 3 conv (32, 64, 96) & gated conv, flattened into & 3 dilated conv (32, 64, 96) \\
 & & a 1.3M-parameter linear layer & \\
Head & 3 FC layers & 4 FC layers & 2 FC layers \\
Parameters & 1.9M & 3.58M & 0.80M \\
Balanced acc., Davis & -- & 0.820 & 0.848 \\
Balanced acc., KIBA & -- & 0.825$^*$ & 0.867 \\
\bottomrule
\end{tabular}
\end{table*}

\paragraph{Drug representation.} We read the drug as its SMILES string, a sequence of characters, truncated or padded to a fixed length $L_D$ (85 on Davis and 100 on KIBA and BindingDB). Padding to a fixed length also hides the true length of the drug from the servers. A sequence is a flat array, and a convolution over it is a linear operation with public filters. This single choice removes all secret-by-secret products from the drug tower (idea~1 above).

\paragraph{Drug embedding.} Each character is mapped to a learned 128-dimensional vector. In plaintext, this is a table lookup. Under MPC, looking up a row of a table at a secret index is expensive. We instead write the drug as a one-hot matrix $O \in \{0, 1\}^{L_D \times V}$, with one row per character and one column per symbol of the vocabulary of size $V = 65$. Row $p$ of $O$ has a single $1$, in the column of the $p$-th character. With the public embedding table $E \in \mathbb{R}^{V \times 128}$, the embedding is the matrix product
\begin{equation}
X^{(0)} = O\,E ,
\label{eq:embed}
\end{equation}
which selects, for each character, its row of $E$. The one-hot matrix is secret and the table is public, so Equation~\eqref{eq:embed} is a secret-by-public product, computed locally with $\pi_{\textsc{MatMul}}$ (Equation~\eqref{eq:matmul}). The drug owner shares $O$ instead of the string, with entries encoded in fixed point as $0$ or $2^{16}$.

Each tower applies three one-dimensional convolutions, each followed by a ReLU. A convolution with kernel size $k$, input $X$ with $C_{\mathrm{in}}$ channels, and public filters $w$ computes, for output channel $c$ and position $p$,
\begin{equation}
Y[c, p] = b_c + \sum_{c'=1}^{C_{\mathrm{in}}} \sum_{j=0}^{k-1} w[c, c', j] \cdot X[c', p + j] .
\label{eq:conv}
\end{equation}
In the drug tower, $X$ is secret and $w$ is public, so the convolution is again a local secret-by-public product, which we compute with $\pi_{\textsc{Conv}}$ (Section~\ref{sec:secure}). The drug tower uses kernel size 4 and only 16, 32, and 48 channels, half of DeepDTA's. Fewer channels mean fewer secret activations, and so fewer ReLU comparisons (idea~2 above).

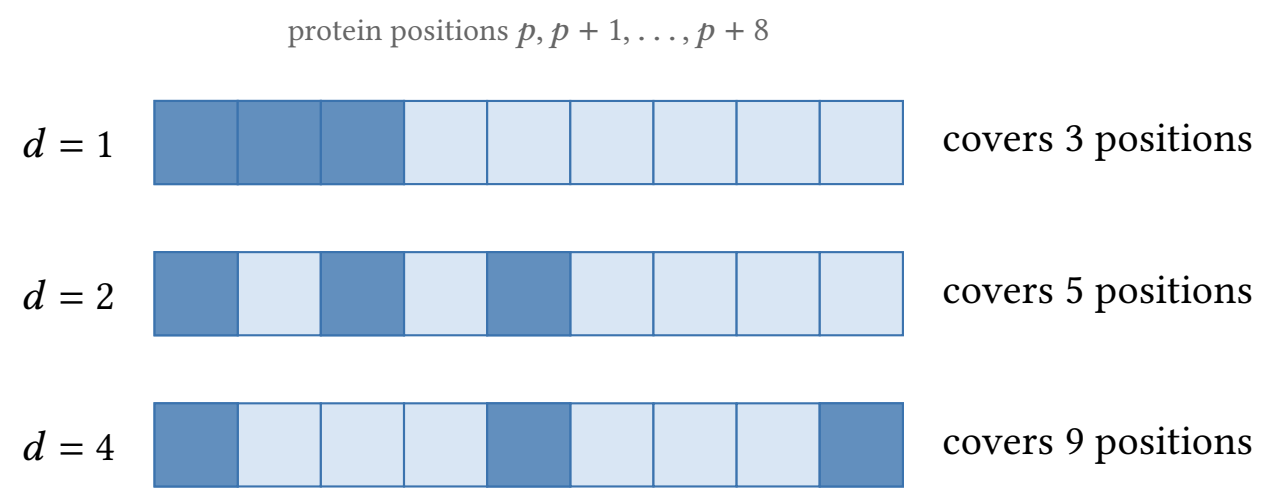
\begin{figure}[t]
\centering
\begin{adjustbox}{max width=\columnwidth}
\begin{tikzpicture}
  \def\W{0.55}
  \foreach \d/\row/\taps in {1/0/{0,1,2}, 2/1/{0,2,4}, 4/2/{0,4,8}} {
     \pgfmathsetmacro{\yR}{-1.0*\row}
     \foreach \i in {0,...,8} {
        \pgfmathsetmacro{\xL}{\i*\W}
        \fill[pubfill, draw=publine] (\xL,\yR) rectangle ++(\W,\W);
     }
     \foreach \t in \taps {
        \pgfmathsetmacro{\xL}{\t*\W}
        \fill[publine!80, draw=publine] (\xL,\yR) rectangle ++(\W,\W);
     }
     \node[left, font=\small] at (-0.15,\yR+0.5*\W) {$d = \d$};
     \pgfmathtruncatemacro{\spanv}{2*\d+1}
     \node[right, font=\small] at (9*\W+0.15,\yR+0.5*\W) {covers $\spanv$ positions};
  }
  \node[font=\scriptsize, text=black!60] at (4.5*\W, 1.0) {protein positions $p, p+1, \ldots, p+8$};
\end{tikzpicture}
\end{adjustbox}
\caption{Dilated convolution with $k = 3$ taps. The same three weights (dark
cells) are spaced $d$ positions apart, so a larger $d$ covers a wider window,
$(k-1)\cdot d + 1$ positions, with no extra weights. In \model, the protein tower
is public, so the wider view adds no secure computation.}
\label{fig:dilation}
\end{figure}

\paragraph{Protein tower.} The protein tower uses kernel size 8 and 32, 64, and 96 channels, and its convolutions are \emph{dilated}~\citep{yu2016multiscale}. A dilated convolution with dilation $d$ spaces its taps $d$ positions apart:
\begin{equation}
Y[c, p] = b_c + \sum_{c'} \sum_{j=0}^{k-1} w[c, c', j] \cdot X[c', p + d \cdot j] .
\label{eq:dilated}
\end{equation}
With the same $k$ weights, a dilated convolution covers $(k-1) \cdot d + 1$ positions instead of $k$ (Figure~\ref{fig:dilation}). With dilations 1, 2, and 4, each output of the protein tower sees 50 amino acids instead of 22. This matters for the kinase proteins in Davis and KIBA, which bind many drugs through a flexible stretch of about 30 amino acids. The most important protein channel of \model responds to exactly such a 29-amino-acid stretch; zeroing it raises the Davis test error (MSE) from 0.291 to 0.322. An undilated tower cannot see this stretch in one piece. In our runs, removing dilation lowered the Davis balanced accuracy from 0.848 to 0.831. Since the protein tower is public, this wider view costs no secure computation (idea~3 above).

After the third convolution, each tower keeps, for each channel, the maximum over all positions. The drug tower thus produces a vector of 48 values, and the protein tower a vector of 96. On the drug side, the maximum over the 76 positions (on Davis) of each of the 48 channels takes $48 \times 75 = 3{,}600$ comparisons in $\lceil \log_2 76 \rceil = 7$ levels of the tree of Section~\ref{sec:fss}, that is, 21 rounds. We also tried mean pooling, which is linear and therefore free, but max pooling was more accurate within the budget.

The head concatenates the drug vector $h_D \in \mathbb{R}^{48}$ and the protein vector $h_T \in \mathbb{R}^{96}$, and applies two fully connected layers with 1024 and 512 units and ReLU, followed by a linear output. We split the first weight matrix at the drug--protein boundary:
\begin{equation}
[\,h_D \mid h_T\,]\, W_1 \;=\; h_D\, W_{1,D} \;+\; h_T\, W_{1,T} .
\label{eq:split}
\end{equation}
The second term depends only on public values, so both servers compute it in plaintext and add it to the secret first term with Equation~\eqref{eq:addc}. The secret product shrinks from 144 to 48 inputs.

Per sample, the secret part of \model performs about 2.6 million multiply-adds, all of them local. It needs $9{,}024$ ReLUs and $3{,}600$ comparisons for max pooling, a total of $12{,}624$ comparisons, which are the only operations that communicate. Appendix~\ref{app:layers} lists the cost of each layer.

\section{MPC on GPUs}
\label{sec:gpu}

GPUs accelerated deep learning, so it is fair to expect that they can also accelerate secure inference, since secure inference performs the same matrix multiplications as plaintext inference, only on shares. As we explain below, this is not straightforward.

A GPU runs thousands of threads in parallel. It is fastest on large, regular workloads in which every thread performs the same operation on different data, such as a large matrix multiplication. Vendor libraries, such as cuBLAS, and specialized units called \emph{tensor cores}, make floating-point matrix products extremely fast, and recent GPUs also support fast products of 8-bit integers. 

A GPU has a memory hierarchy. Each thread has a few fast \emph{registers}, groups of threads share a small fast \emph{shared memory}, and all threads can access a large but slow \emph{global memory}. Launching a GPU function, called a \emph{kernel}, has a fixed overhead of several microseconds. A workload that consists of many small kernels, or that moves data in and out of global memory at every step, cannot use the GPU well.

\paragraph{Problem 1: integer matrix multiplication.} Our shares are 64-bit integers modulo $2^{64}$, and the linear layers are matrix products of such integers. GPUs do not offer a fast 64-bit integer matrix product. Tensor cores and cuBLAS are built for floating-point numbers and for small integers. In PyTorch, for example, a matrix product of two 64-bit integer tensors on a CUDA device is simply not implemented. A naive workaround, such as computing in 64-bit floating point, loses precision, because a 64-bit float has only 53 bits for the digits of a number, and represents integers exactly only up to $2^{53}$. For example, $2^{53} + 1$ is rounded to $2^{53}$. Shares are uniformly random 64-bit numbers, so almost every share is larger than $2^{53}$, and even a single product of two shares has up to 128 bits before the reduction modulo $2^{64}$. A single rounding error in a share changes the secret, which makes the result wrong.

\paragraph{Problem 2: tree-based comparison.} An FSS comparison is a walk down a 63-level tree, with AES evaluations at every level, and each level depends on the previous one. A single comparison is therefore a chain of 63 steps that cannot run in parallel, and the only parallelism is across the thousands of independent comparisons. This workload is the opposite of a matrix multiplication. It does not use tensor cores, and if each level is implemented as a separate sequence of tensor operations, then each level launches several kernels and writes its intermediate seeds to global memory, only to read them back at the next level. For thousands of comparisons per sample and 63 levels each, this overhead dominates.

\paragraph{Problem 3: communication.} A GPU reduces $T_{\mathrm{comp}}$ in Equation~\eqref{eq:cost}, but not the rounds $R$ or the bytes $S$. \emph{Batching} helps with the rounds. If $B$ samples are processed together, all of them share the same rounds, so the per-sample time becomes
\begin{equation}
\frac{T_{\mathrm{online}}}{B} \;\approx\; \frac{T_{\mathrm{comp}}}{B}
\;+\; \frac{R \cdot \tau}{B} \;+\; \frac{S_{\mathrm{sample}}}{\mathrm{BW}} ,
\label{eq:batch}
\end{equation}
where $S_{\mathrm{sample}}$ is the number of bytes per sample. The first two terms shrink as the batch grows. For example, if a single sample needs 37 rounds and each round costs a latency of $\tau = 1$\,ms, the sample spends 37\,ms waiting on the network, but a batch of 1,000 samples shares these 37\,ms, which costs only 0.037\,ms per sample. The last term does not shrink, because every sample sends its own masked values. This term is a floor that no GPU and no batch size can cross. Only a protocol that sends fewer bytes, or a model that needs fewer comparisons, can lower it.

\paragraph{Problem 4: memory.} Offline material, such as DCF keys and Beaver triples, must be in GPU memory during the online phase. A DCF key holds one correction word for each of the 63 levels, more than a kilobyte per comparison. With 12,624 comparisons per sample in \model, the keys alone take more than 10\,MB per sample and per server. For an input set of 100,000 samples, this is more than a terabyte, while a GPU has tens of gigabytes of memory. The material for a large input set therefore does not fit in memory at once.

Our solution addresses each of these problems, and builds on prior work where it can. For Problem 1, we use the limb decomposition of CryptGPU~\citep{tan2021cryptgpu}. For Problem 2, we wrote our own fused CUDA kernel for DCF evaluation, whose memory layout follows Orca~\citep{jawalkar2024orca}. Where it makes the difference is that each comparison walks all 63 levels in a single kernel launch, without writing intermediate values to global memory (Section~\ref{sec:secure}). This kernel makes the whole secure inference 2.9 to 4.3 times faster than a tree walk written with PyTorch tensor operations (Section~\ref{sec:results}). For Problem 3, we combine a protocol that sends few bytes, FSS~\citep{boyle2021function}, with our MPC-friendly model, which needs few comparisons (Section~\ref{sec:model}). For Problem 4, we split large input sets into batches that fit in memory. Our contribution is not any single technique, but their combination with the model design, which together make secure inference of a DTI model fast on GPUs (Section~\ref{sec:secure}).

\section{Secure Inference of \model on GPUs}
\label{sec:secure}

\begin{figure*}[t]
\centering
\resizebox{\textwidth}{!}{%
\begin{tikzpicture}[font=\Large, >={Stealth[length=2.2mm]},
  L/.style={draw=locline, line width=1.2pt, fill=secfill, rounded corners=2pt, minimum height=1.3cm, minimum width=1.45cm, align=center, inner sep=3pt},
  C/.style={draw=comline, line width=1.2pt, fill=secfill, rounded corners=2pt, minimum height=1.3cm, minimum width=1.45cm, align=center, inner sep=3pt},
  r/.style={font=\large, text=black!65}]
\node[draw=secline, fill=secfill, rounded corners=2pt, minimum height=1.0cm, align=center, inner sep=3pt] (in) at (0,0) {one-hot\\$\share{O}$};
\node[L] (e)  at (1.8,0)  {embed\\$\share{O}E$};
\node[L] (c1) at (3.5,0)  {conv 1};
\node[C] (r1) at (5.2,0)  {ReLU};
\node[L] (c2) at (6.9,0)  {conv 2};
\node[C] (r2) at (8.6,0)  {ReLU};
\node[L] (c3) at (10.3,0) {conv 3};
\node[C] (r3) at (12.0,0) {ReLU};
\node[C] (mp) at (13.7,0) {max\\pool};
\node[L] (f1) at (15.4,0) {FC 1};
\node[C] (r4) at (17.1,0) {ReLU};
\node[L] (f2) at (18.8,0) {FC 2};
\node[C] (r5) at (20.5,0) {ReLU};
\node[L] (f3) at (22.2,0) {output};
\node[draw=secline, fill=secfill, rounded corners=2pt, minimum height=1.0cm, align=center, inner sep=3pt] (out) at (23.9,0) {$\share{\hat y}$};
\foreach \a/\b in {in/e, e/c1, c1/r1, r1/c2, c2/r2, r2/c3, c3/r3, r3/mp, mp/f1, f1/r4, r4/f2, f2/r5, r5/f3, f3/out}
  \draw[->, black!55] (\a) -- (\b);
\foreach \n in {e, c1, c2, c3, f1, f2, f3}
  \node[r, below=2pt of \n] {0 rounds};
\foreach \n/\t in {r1/3, r2/3, r3/3, mp/21, r4/3, r5/3}
  \node[r, below=2pt of \n, text=comline] {\t\ rounds};
\node[draw=publine, fill=pubfill, rounded corners=2pt, minimum height=0.9cm, text width=13.2cm, align=center, inner sep=4pt] (prot) at (7.6,2.5)
  {protein tower in plaintext: embed, 3 dilated convolutions with ReLU, max pool\\(separate GPU stream, overlaps the waiting time of the drug tower)};
\draw[->, publine, thick] (prot.east) -| node[pos=0.8, right, r, text=publine] {$h_T W_{1,T}$} (f1.north);
\node[L, minimum width=0.5cm, minimum height=0.35cm] (lg1) at (1.0,-1.6) {};
\node[anchor=west, r, text=black] at (lg1.east) {local, no communication};
\node[C, minimum width=0.5cm, minimum height=0.35cm] (lg2) at (7.0,-1.6) {};
\node[anchor=west, r, text=black] at (lg2.east) {communicates (FSS comparisons)};
\node[anchor=west, r, text=black] at (14.2,-1.6) {plus 1 round to deliver the prediction, 37 rounds in total};
\end{tikzpicture}%
}
\caption{The secure evaluation of \model. Orange marks values that depend on
the drug, and blue marks public values. Layers with a green border are linear
with public weights and run locally on each server's GPU. Layers with a red
border are comparisons evaluated with FSS, and are the only layers that
communicate. The protein tower (blue) runs in plaintext and joins at the first fully connected layer through the
split of Equation~\eqref{eq:split}. The rounds do not depend on the batch size.}
\label{fig:pipeline}
\end{figure*}
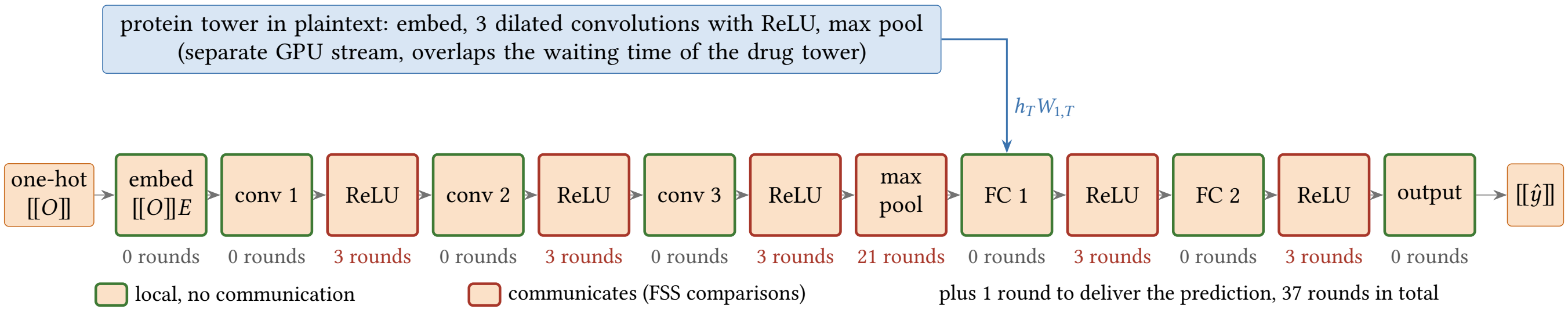

Figure~\ref{fig:pipeline} shows the secure evaluation of \model. The protein tower runs in plaintext. Everything that depends on the drug is held as additive shares over $\mathbb{Z}_{2^{64}}$, with 16 fractional bits. Linear layers (green border) are local, and only the ReLU and max-pooling layers (red border) communicate. In this section, we first describe the whole protocol, and then how each part runs on the GPU.

Algorithm~\ref{alg:slimdta} lists the protocol $\pi_{\textsc{SlimDTA}}$ that each server runs on a batch, built from the protocols of Table~\ref{tab:cost}. Line~1 runs the protein tower in plaintext, on a separate CUDA stream, so that it computes while the drug tower waits for the network. Line~2 computes the embedding of the shared one-hot matrix (Equation~\eqref{eq:embed}). Lines~3 to~5 apply the three convolutions of the drug tower, each followed by a ReLU, and line~6 takes the maximum of each channel. Line~7 computes the first fully connected layer with the split of Equation~\eqref{eq:split}. The secret part is a call to $\pi_{\textsc{MatMul}}$, and the public part $h_T W_{1,T} + b_1$ is added by $P_0$ alone (Equation~\eqref{eq:addc}). Lines~8 to~10 finish the head, and line~11 sends each server's share of the prediction to the drug owner. The comments give the rounds of each line. They add up to $3 \cdot 3 + 21 + 3 + 3 + 1 = 37$ rounds, independent of the batch size.

\begin{algorithm}[!htb]
\caption{$\pi_{\textsc{SlimDTA}}(\share{O}, T)$, run by each server $P_i$ on a batch.}
\label{alg:slimdta}
\begin{algorithmic}[1]
\State $h_T \gets$ protein tower on $T$, in plaintext \Comment{0 rounds}
\State $\share{X} \gets \pi_{\textsc{MatMul}}(\share{O}, E)$ \Comment{0 rounds}
\For{$l = 1, 2, 3$}
  \State $\share{X} \gets \pi_{\textsc{ReLU}}\big(\pi_{\textsc{Conv}}(\share{X}, w_l, b_l)\big)$ \Comment{3 rounds}
\EndFor
\State $\share{h_D} \gets \pi_{\textsc{MaxPool}}(\share{X})$ per channel \Comment{21 rounds}
\State $\share{z} \gets \pi_{\textsc{MatMul}}(\share{h_D}, W_{1,D}) + (h_T W_{1,T} + b_1)$ \Comment{$P_0$ adds}
\State $\share{z} \gets \pi_{\textsc{ReLU}}(\share{z})$ \Comment{3 rounds}
\State $\share{z} \gets \pi_{\textsc{ReLU}}\big(\pi_{\textsc{FC}}(\share{z}, W_2, b_2)\big)$ \Comment{3 rounds}
\State $\share{\hat y} \gets \pi_{\textsc{FC}}(\share{z}, W_3, b_3)$
\State send the share of $\hat y$ to the drug owner \Comment{1 round}
\end{algorithmic}
\end{algorithm}

\subsection{Linear Layers}

We compute each convolution, $\pi_{\textsc{Conv}}$, as a single matrix product, by the standard \emph{im2col} transformation.\footnote{im2col (image to column) copies every window of the input that a filter covers into one row of a matrix. For example, for a single-channel input $(x_1, x_2, x_3, x_4)$ and a filter of size 2, the rows are $(x_1, x_2)$, $(x_2, x_3)$, and $(x_3, x_4)$. Multiplying this matrix by the filter weights gives all outputs of the convolution at once. With $C$ input channels and kernel size $k$, each row has $C k$ entries.} The input windows of Equation~\eqref{eq:conv} are rearranged into the rows of a matrix, so that the convolution becomes the product of this matrix with the public filters. Rearranging a shared matrix is local, since each server rearranges its own share. The embedding (Equation~\eqref{eq:embed}), the convolutions, and the fully connected layers of the head, $\pi_{\textsc{FC}}$, are therefore all calls to $\pi_{\textsc{MatMul}}$, a product of a secret matrix and a public matrix. We process the whole batch in a single product, which makes the matrices large enough to use the GPU well.

\begin{figure}[!htb]
\centering
\begin{adjustbox}{max width=\columnwidth}
\begin{tikzpicture}[font=\small]
  \node[draw=secline, fill=secfill, minimum width=3.6cm, minimum height=0.6cm] (a) at (1.8,0.55) {share $a$ (64 bits)};
  \foreach \i/\x in {3/0.45,2/1.35,1/2.25,0/3.15}
    \node[draw=secline, fill=secfill, minimum width=0.9cm, minimum height=0.6cm] (a\i) at (\x,-0.45) {$a_\i$};
  \draw[->, >={Stealth[length=2mm]}] (a.south) -- (1.8,-0.15);
  \node[font=\footnotesize, text=black!65, align=center] at (1.8,-1.25) {four 16-bit limbs\\$a = a_0 + 2^{16} a_1 + 2^{32} a_2 + 2^{48} a_3$};
  \draw[fill=secfill, draw=locline, line width=0.9pt] (0.05,-2.0) rectangle ++(0.3,-0.3);
  \node[anchor=west, font=\footnotesize] at (0.4,-2.15) {kept, one GPU product each};
  \draw[fill=black!7, draw=black!30] (0.05,-2.45) rectangle ++(0.3,-0.3);
  \node[anchor=west, font=\footnotesize] at (0.4,-2.6) {multiple of $2^{64}$, dropped};
  \foreach \i in {0,...,3}{
    \foreach \j in {0,...,3}{
      \pgfmathtruncatemacro{\s}{\i+\j}
      \ifnum\s<4
        \draw[fill=secfill, draw=locline, line width=0.9pt] ({4.6+0.8*\j},{0.8-0.8*\i}) rectangle ++(0.8,-0.8);
        \node[font=\small] at ({5.0+0.8*\j},{0.4-0.8*\i}) {$a_\i b_\j$};
      \else
        \draw[fill=black!7, draw=black!30] ({4.6+0.8*\j},{0.8-0.8*\i}) rectangle ++(0.8,-0.8);
      \fi
    }
  }
  \foreach \j in {0,...,3} \node[font=\small] at ({5.0+0.8*\j},1.05) {$b_\j$};
  \foreach \i in {0,...,3} \node[font=\small] at (4.32,{0.4-0.8*\i}) {$a_\i$};
\end{tikzpicture}
\end{adjustbox}
\caption{Computing a product modulo $2^{64}$ with limbs. Each 64-bit value is
split into four 16-bit limbs (left). Of the 16 limb products, only the 10 with
$i + j < 4$ contribute modulo $2^{64}$ (right). Each kept product is an exact
floating-point matrix product on the GPU.}
\label{fig:limbs}
\end{figure}
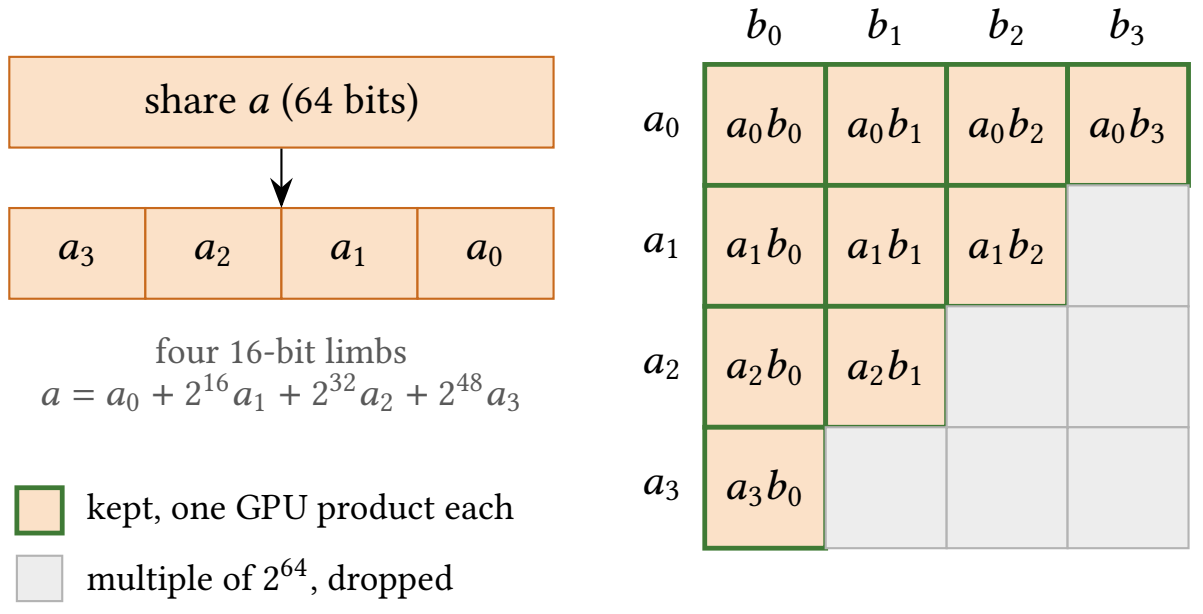

\paragraph{Integer matrix multiplication.} To compute these products exactly modulo $2^{64}$ on a GPU, we use the idea of long multiplication by hand, in which each number is split into digits, the digits are multiplied pairwise, and the partial products are shifted and added. We split each 64-bit entry $a$ of a share into four 16-bit \emph{limbs},
\begin{equation}
a = \sum_{i=0}^{3} a_i \, 2^{16 i}, \qquad 0 \le a_i < 2^{16} ,
\label{eq:limb}
\end{equation}
and do the same for each public weight $b$. Then
\begin{equation}
a \cdot b \;\equiv\; \sum_{i + j < 4} (a_i\, b_j) \, 2^{16 (i+j)} \pmod{2^{64}} .
\label{eq:limbprod}
\end{equation}
Every term with $i + j \ge 4$ is a multiple of $2^{64}$ and vanishes, so only 10 of the 16 limb products are needed (Figure~\ref{fig:limbs}). For matrices, each term $a_i b_j$ becomes a product of two limb matrices, which we compute as a 64-bit floating-point matrix product with cuBLAS. Each entry of such a product is a sum of $K$ terms, each smaller than $2^{32}$, where $K$ is the inner dimension of the product. The result is exact as long as it stays below $2^{53}$, which holds for $K < 2^{21}$. The largest inner dimension in \model is 1,024, far below this limit. Appendix~\ref{app:limb} gives the details. This decomposition follows CryptGPU~\citep{tan2021cryptgpu}. On GPUs with integer tensor cores, an optional variant splits each value into eight 8-bit limbs and computes the surviving limb products as 8-bit integer matrix products on the tensor cores.

After each product, the result carries 32 fractional bits, and each server truncates its share locally by 16 bits with $\pi_{\textsc{Trunc}}$ (Appendix~\ref{app:trunc}). No communication is needed. Algorithm~\ref{alg:matmul} summarizes $\pi_{\textsc{MatMul}}$, which each server runs on its own share $X_i$. Line~1 splits the share and the public weights into 16-bit limbs (Equation~\eqref{eq:limb}). Lines~3 to~6 compute the 10 limb products that survive modulo $2^{64}$, each as one floating-point matrix product with cuBLAS, which is exact because $K < 2^{21}$. Each product is converted back to a 64-bit integer, shifted by $16(a + c)$ bits, and added to the result. Line~7 truncates the result locally.

\begin{algorithm}[t]
\caption{$\pi_{\textsc{MatMul}}(\share{X}, W)$, run by $P_i$ on its share $X_i$.}
\label{alg:matmul}
\begin{algorithmic}[1]
\State split $X_i$ and $W$ into 16-bit limbs $X_i^{(a)}$ and $W^{(c)}$ \Comment{Eq.~\eqref{eq:limb}}
\State $Y_i \gets 0$
\For{all $(a, c)$ with $a + c < 4$} \Comment{10 pairs}
  \State $Z \gets X_i^{(a)} W^{(c)}$ in FP64 with cuBLAS \Comment{exact}
  \State $Y_i \gets Y_i + 2^{16(a+c)} \cdot Z \bmod 2^{64}$
\EndFor
\State \Return $\pi_{\textsc{Trunc}}(Y_i)$ \Comment{local}
\end{algorithmic}
\end{algorithm}

\subsection{Non-Linear Layers}

Each ReLU layer runs $\pi_{\textsc{ReLU}}$ (Algorithm~\ref{alg:relu}) on all its elements at once. All elements of the layer, across the whole batch, are masked and opened in one round (line~1), evaluated with the DCF in one kernel launch (line~2), and converted and multiplied in two more rounds (lines~4 and~5). A ReLU layer therefore costs three rounds, whatever its size and the batch size.

Max-pooling runs $\pi_{\textsc{MaxPool}}$, the tree of Section~\ref{sec:fss}, where all comparisons of a level, in all channels and all samples, form one batched step of three rounds. When a level has an odd number of values, we carry the unpaired value forward to the next level, instead of padding the channel to a power of two. For the 76 values of a drug channel, padding to 128 would need 127 comparisons per channel instead of 75, so carrying the value forward saves 17\% of all comparisons, and of the communication.

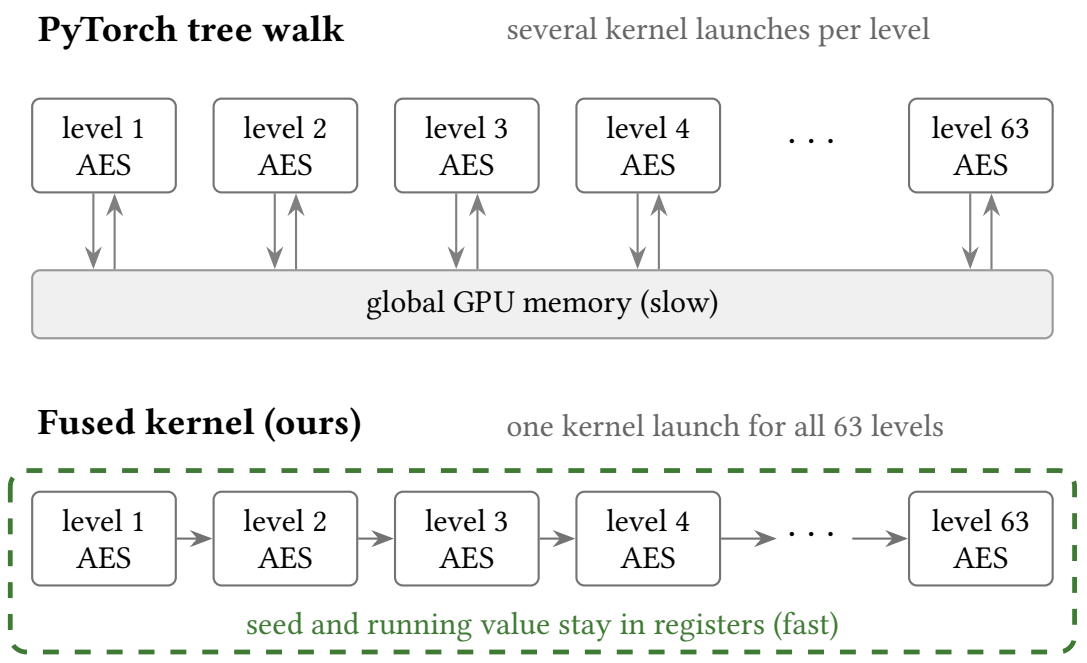
\begin{figure}[!htb]
\centering
\begin{adjustbox}{max width=\columnwidth}
\begin{tikzpicture}[font=\footnotesize, >={Stealth[length=1.5mm]},
  lv/.style={draw=black!55, fill=white, rounded corners=1.5pt, minimum width=0.95cm, minimum height=0.62cm, inner sep=1pt, align=center, font=\scriptsize},
  hd/.style={font=\footnotesize\bfseries, anchor=west}]
\node[hd] at (-0.55,1.15) {PyTorch tree walk};
\node[font=\scriptsize, text=black!60, anchor=west] at (2.55,1.15) {several kernel launches per level};
\foreach \t/\x in {1/0.0,2/1.2,3/2.4,4/3.6} \node[lv] (a\t) at (\x,0.4) {level \t\\AES};
\node at (4.7,0.4) {$\cdots$};
\node[lv] (a63) at (5.8,0.4) {level 63\\AES};
\node[draw=black!40, fill=black!6, rounded corners=1.5pt, minimum width=6.75cm, minimum height=0.45cm, font=\scriptsize] (gm) at (2.9,-0.65) {global GPU memory (slow)};
\foreach \t in {1,2,3,4,63} {\draw[->, black!50] ([xshift=-2pt]a\t.south) -- ([xshift=-2pt]a\t.south |- gm.north); \draw[<-, black!50] ([xshift=2pt]a\t.south) -- ([xshift=2pt]a\t.south |- gm.north);}
\node[hd] at (-0.55,-1.45) {Fused kernel (ours)};
\node[font=\scriptsize, text=black!60, anchor=west] at (2.55,-1.45) {one kernel launch for all 63 levels};
\foreach \t/\x in {1/0.0,2/1.2,3/2.4,4/3.6} \node[lv] (b\t) at (\x,-2.2) {level \t\\AES};
\node at (4.7,-2.2) {$\cdots$};
\node[lv] (b63) at (5.8,-2.2) {level 63\\AES};
\foreach \p/\q in {b1/b2,b2/b3,b3/b4} \draw[->, black!55] (\p) -- (\q);
\draw[->, black!55] (b4) -- (4.45,-2.2); \draw[->, black!55] (4.95,-2.2) -- (b63);
\draw[locline, dashed, line width=0.9pt, rounded corners=3pt] (-0.62,-1.75) rectangle (6.42,-2.95);
\node[font=\scriptsize, text=locline] at (2.9,-2.78) {seed and running value stay in registers (fast)};
\end{tikzpicture}
\end{adjustbox}
\caption{Two ways to evaluate a DCF key on the GPU. A tree walk written with
PyTorch tensor operations launches several kernels per level and writes the
intermediate seeds of every comparison to global memory, only to read them back
at the next level. Our fused kernel walks all 63 levels in one launch, with the
seed and the running value in registers. In both cases, each GPU thread
handles one comparison, and all comparisons of a layer run in parallel.}
\label{fig:kernel}
\end{figure}

\paragraph{DCF evaluation kernel.} Most of the online computation is the DCF evaluation of line~2 of Algorithm~\ref{alg:relu}, with 63 tree levels for each of 12,624 comparisons per sample and two AES calls per level. Our first implementation expressed each level as a sequence of PyTorch tensor operations. This worked, but each level launched several kernels and wrote all intermediate seeds to global memory, which is the inefficiency described in Section~\ref{sec:gpu}. We replaced it with a single fused CUDA kernel (Figure~\ref{fig:kernel}). Each GPU thread evaluates one comparison and walks all 63 levels of its tree in one launch. The 128-bit seed and the running value stay in registers for the whole walk, so no intermediate values go to global memory. The AES lookup tables sit in shared memory, which is fast. The correction words are stored level by level, so that threads that evaluate neighboring comparisons read neighboring memory addresses at the same time, which the GPU can serve in a single memory transaction. This layout follows Orca~\citep{jawalkar2024orca}. The kernel returns the same shares as the PyTorch implementation, bit for bit.

\subsection{Batching and Chunking}

An input set can contain from hundreds to hundreds of thousands of samples (Section~\ref{sec:setting}). We split it into \emph{batches} of $B$ samples, and run the protocol on one batch at a time. We call this splitting \emph{chunking}. The batch size $B$ trades off two of the problems of Section~\ref{sec:gpu}.

A larger batch reduces the cost of rounds (Problem~3). The $B$ samples of a batch go through every layer together. In each round, each server sends a single message that holds the masked values of all $B$ samples, instead of one message per sample. The 37 rounds are therefore paid once per batch, and their cost per sample shrinks as $B$ grows (Equation~\eqref{eq:batch}). A larger batch also gives the GPU larger matrix products and more comparisons to evaluate in parallel.

A smaller batch needs less memory (Problem~4). The offline material of a batch, that is, the masks, DCF keys, and Beaver triples that the dealer prepares for it, grows linearly with $B$. The servers load the material of one batch, run the protocol, and discard the material before the next batch. The memory needed therefore depends on $B$ and not on the size of the input set, which lets us process input sets of any size on a fixed GPU. The predictions do not depend on $B$, up to fixed-point error. The material does not depend on the drug, so the dealer generates it offline, and the online time starts once the material of a batch is in memory.

The protein tower (line~1 of Algorithm~\ref{alg:slimdta}) runs on a separate CUDA stream, so that its computation overlaps the waiting time of the drug tower. It is still part of the timed online phase.

\subsection{Security}

Under the threat model of Section~\ref{sec:setting}, the protocol reveals nothing about the drug to either server. A formal proof follows from the security of the individual building blocks and the composition of semi-honest protocols~\citep{canetti2000security, boyle2021function}. The servers learn only public information, namely the protein, the weights, the padded length, and the batch size.

\section{Results}
\label{sec:results}

\subsection{Competition Results}

The organizers evaluated the solutions on two workstations with NVIDIA H100 GPUs, one per server, connected by a network of more than 1\,Gbps, with batch size 64. They timed three tests: 1,000 and 10,000 samples from BindingDB~\citep{liu2007bindingdb}, and 1,000 samples from Davis~\citep{davis2011comprehensive}. Table~\ref{tab:competition} shows the results for the four finalist teams. Our solution was the fastest on all three tests. Dividing the total time reported by the organizers by the number of samples, it took 9.2\,ms per sample on BindingDB and 8.1\,ms per sample on Davis. These are averages over the whole run, which processed the samples in batches of 64, and not a measurement of the latency of a single prediction. Compared with the next fastest solution on each test, it was 1.52 times faster on 1,000 BindingDB samples, 1.16 times faster on 10,000 BindingDB samples, and 1.17 times faster on Davis.

\begin{table*}[t]
\centering
\caption{Secure inference time in the iDASH 2026 Track~III evaluation, as
reported by the organizers. Two H100 workstations, a network of more than
1\,Gbps, and batch size 64.}
\label{tab:competition}
\small
\begin{tabular}{@{}lrrr@{}}
\toprule
Team & BindingDB (1,000 samples) & BindingDB (10,000 samples) & Davis (1,000 samples) \\
\midrule
\textbf{PPMLHuskies (ours)} & \textbf{9.2\,s} & \textbf{91.5\,s} & \textbf{8.1\,s} \\
Team A & 14.4\,s & 106.1\,s & 9.5\,s \\
Team B & 14.0\,s & 164.6\,s & 12.7\,s \\
Team C & 2{,}486.0\,s & 25{,}463.5\,s & 2{,}518.9\,s \\
\midrule
Ours, per sample & 9.2\,ms & 9.2\,ms & 8.1\,ms \\
Ours, speedup over the next fastest & 1.52$\times$ & 1.16$\times$ & 1.17$\times$ \\
\bottomrule
\end{tabular}
\end{table*}

\subsection{Experimental Setup}

For our own experiments, we used one workstation with two NVIDIA RTX 6000 Ada GPUs with 48\,GB of memory each. We ran the two servers as two separate processes, each with its own GPU, that exchange real messages over sockets. To reproduce a realistic network, we limited the link between them to 1\,Gbps with an added delay of 0.25\,ms in each direction, using the Linux traffic control tool (\texttt{tc netem}). This is the lowest bandwidth the competition allowed. Since both servers run on one machine, the throttle applies to the loopback interface, so the traffic of both servers, in both directions, shares a single 1\,Gbps link. Between two machines with a full-duplex link, each direction would have its own 1\,Gbps. Our communication times are therefore conservative, by up to a factor of two. We measure the online time per sample, from the moment the servers hold the input shares until the shares of the predictions are delivered, including the plaintext protein tower. In our experiments, this last round opens the predictions so that we can check their error. In deployment, the servers instead send their shares to the drug owner. We synchronize the GPU before reading the timer, and we load the CUDA kernel before timing starts. The numbers of rounds and bytes are counted by the framework, and are exact. Unless stated otherwise, we use the Davis test set.

We compare three ways of evaluating the comparisons. All three use the same model, the same input sharing, and the same local linear layers.
\begin{itemize}[leftmargin=1.4em, itemsep=1pt]
\item \textbf{FSS + CUDA kernel} (ours): FSS comparisons, with the fused DCF kernel of Section~\ref{sec:secure}.
\item \textbf{FSS + PyTorch}: the same FSS comparisons, with the DCF tree walk written as PyTorch tensor operations.
\item \textbf{Bit decomposition}: comparisons through a ripple-carry adder on bit shares, using edaBits~\citep{escudero2020improved}.
\end{itemize}

\vspace{0.5em}
\noindent\textbf{Accuracy.} In plaintext, the \model checkpoint that we submitted reaches a balanced accuracy of 0.848 on Davis, against 0.820 for DeepDTAGen, and 0.867 against 0.825 on KIBA (Table~\ref{tab:compare}). This checkpoint is our best training run. Over training seeds, the Davis balanced accuracy of \model averages about 0.82, so \model is on par with DeepDTAGen, with 4.5 times fewer parameters. The secure predictions are close to the plaintext predictions of \model. The only source of error is fixed-point arithmetic, and the mean relative error of the secure predictions is 0.015\%, more than a hundred times below the 2\% budget. On the full Davis test set, the secure predictions give the same balanced accuracy as the plaintext model, 0.848.

\begin{figure}[t]
\centering
\begin{tikzpicture}
\begin{axis}[
  width=\columnwidth, height=5.4cm,
  xmode=log, log basis x=2,
  ymode=log,
  xtick={16,64,256,512,1024},
  xticklabels={16,64,256,512,1024},
  ytick={10,30,100,300},
  yticklabels={10,30,100,300},
  xlabel={batch size},
  ylabel={online time per sample (ms)},
  ymin=6, ymax=500,
  grid=major, grid style={black!10},
  legend style={at={(0.5,-0.30)}, anchor=north, legend columns=2, font=\footnotesize, draw=none, /tikz/every even column/.append style={column sep=6pt}},
  legend cell align=left,
]
\addplot[mark=*, very thick, black] coordinates {(16,67.8) (64,20.9) (256,11.5) (512,10.0) (1024,9.5)};
\addlegendentry{FSS + CUDA kernel (ours)}
\addplot[mark=square*, thick, black!55] coordinates {(16,293.4) (64,72.8) (256,33.7) (512,32.1) (1024,34.7)};
\addlegendentry{FSS + PyTorch}
\addplot[mark=triangle*, thick, black!35, densely dotted, mark options={solid}] coordinates {(16,386.3) (64,252.0) (256,223.4) (512,219.0)};
\addlegendentry{bit decomposition (edaBits)}
\addplot[dashed, black!50, domain=16:1024, samples=2] {8.6};
\addlegendentry{measured communication time}
\end{axis}
\end{tikzpicture}
\caption{Online time per sample for real two-party inference over a 1\,Gbps link
with 0.25\,ms delay, on the Davis model (log scales). Our solution approaches the
communication time, shown dashed, which no amount of GPU computation can reduce.}
\label{fig:sweep}
\end{figure}
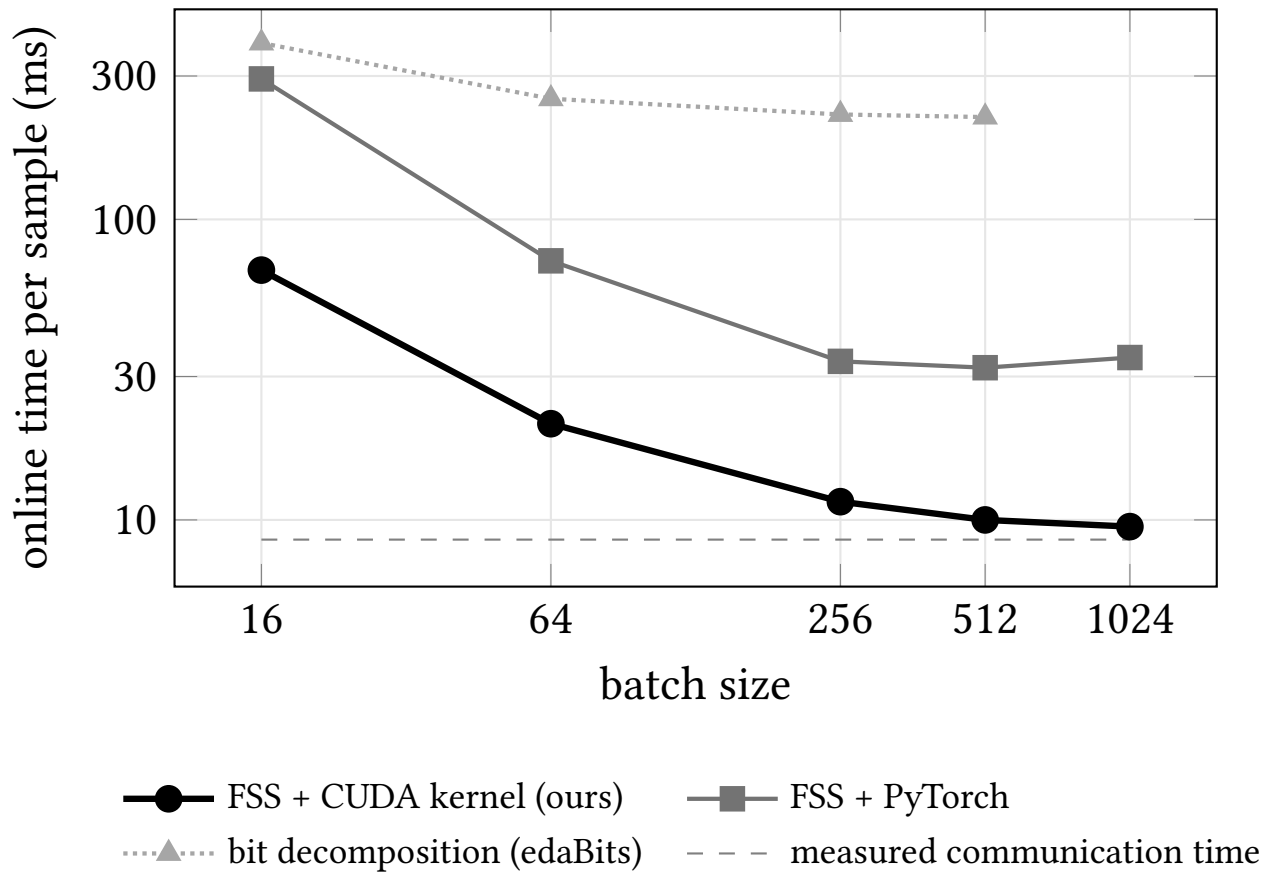

\vspace{0.5em}
\noindent\textbf{Effect of Batch Size.}\label{sec:results-batch} Figure~\ref{fig:sweep} and Table~\ref{tab:sweep} show the online time per sample for batch sizes from 16 to 1,024. Our solution reaches 10.0\,ms per sample at batch size 512 and 9.5\,ms at 1,024. Three observations stand out.

First, FSS is much faster than bit decomposition at every batch size. At batch size 512, it is 21.9 times faster (10.0\,ms against 219.0\,ms). The reason is the number of bytes. Bit decomposition sends 13.2\,MB per sample, against 0.505\,MB for FSS, and on a 1\,Gbps link these bytes dominate. Batching cannot remove this gap, because, as Equation~\eqref{eq:batch} shows, the cost of bytes per sample does not shrink with the batch size. Bit decomposition also ran out of memory at batch size 1,024.

Second, the fused CUDA kernel makes the whole secure inference 2.9 to 4.3 times faster than the PyTorch tree walk. At batch size 256, it reduces the computation time for a batch from 6.48\,s to 0.76\,s, a factor of 8.5. The PyTorch version even becomes slightly slower per sample beyond batch size 512, while the fused kernel keeps improving.

Third, the time per sample levels off. Beyond batch size 512, a larger batch barely helps. The breakdown below explains why.

\begin{table*}[t]
\centering
\caption{Online time per sample on Davis, real two-party inference over a
1\,Gbps link. The right two columns split our time into communication and
computation.}
\label{tab:sweep}
\small
\begin{tabular}{@{}rrrrrr@{}}
\toprule
 & \multicolumn{3}{c}{Online time per sample (ms)} & \multicolumn{2}{c}{Ours, split (ms)} \\
\cmidrule(lr){2-4} \cmidrule(l){5-6}
Batch & FSS + CUDA (ours) & FSS + PyTorch & Bit decomposition & Communication & Computation \\
\midrule
16    & 67.8 & 293.4 & 386.3 & 10.6 & 56.9 \\
64    & 20.9 &  72.8 & 252.0 &  8.9 & 12.0 \\
256   & 11.5 &  33.7 & 223.4 &  8.6 &  3.0 \\
512   & \textbf{10.0} &  32.1 & 219.0 &  8.5 &  1.5 \\
1,024 &  9.5 &  34.7 & out of memory & 8.6 & 0.9 \\
\bottomrule
\end{tabular}
\end{table*}

\begin{figure}[t]
\centering
\begin{tikzpicture}
\begin{axis}[
  ybar stacked, bar width=14pt,
  width=\columnwidth, height=5.4cm,
  symbolic x coords={16,64,256,512,1024},
  xtick=data,
  xlabel={batch size},
  ylabel={time per sample (ms)},
  ymin=0, ymax=72,
  enlarge x limits=0.15,
  legend style={at={(0.5,-0.30)}, anchor=north, legend columns=2, font=\footnotesize, draw=none, /tikz/every even column/.append style={column sep=6pt}},
  legend cell align=left,
]
\addplot[fill=comfill, draw=comline] coordinates {(16,10.6) (64,8.9) (256,8.6) (512,8.5) (1024,8.6)};
\addlegendentry{communication}
\addplot[fill=locfill, draw=locline] coordinates {(16,56.9) (64,12.0) (256,3.0) (512,1.5) (1024,0.93)};
\addlegendentry{computation}
\end{axis}
\end{tikzpicture}
\caption{Where our online time goes. Computation (green) shrinks as the batch
grows. Communication (red) stays at about 8.6\,ms per sample, the time for both
servers to send 0.505\,MB per sample each over the shared 1\,Gbps link.}
\label{fig:breakdown}
\end{figure}
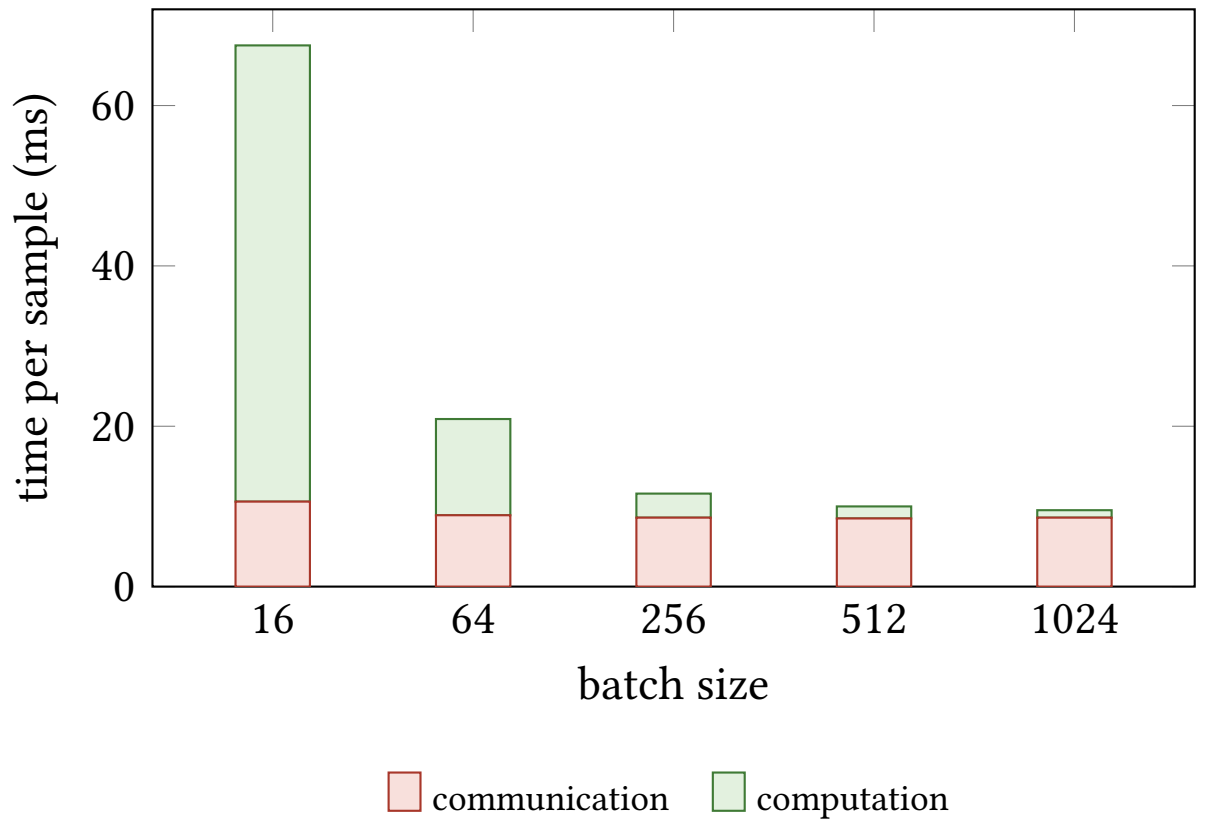

\vspace{0.5em}
\noindent\textbf{Runtime Breakdown.} Figure~\ref{fig:breakdown} splits our online time into communication and computation. The computation per sample falls quickly with the batch size, from 56.9\,ms at batch size 16 to 0.9\,ms at 1,024, because the GPU processes larger batches more efficiently. The communication per sample, in contrast, stays flat at about 8.6\,ms from batch size 64 onward. This is the floor of Equation~\eqref{eq:batch}. Once the rounds are amortized, what remains is the time to send 0.505\,MB per sample from each server. On our shared loopback link, the two servers together send about 1.01\,MB per sample, which takes about 8.1\,ms at 1\,Gbps. The remaining 0.5\,ms is protocol and network overhead. At batch size 1,024, our total of 9.5\,ms is only 0.9\,ms above this floor. Further speedups therefore require sending fewer bytes, not faster GPUs. This is also why the competition numbers, measured on H100 GPUs on two machines with a link faster than 1\,Gbps, are lower than ours at the same batch size.

\vspace{0.5em}
\noindent\textbf{Ablation Study.} Table~\ref{tab:ablation} shows how the setting, the model, and the protocol each reduce the communication. The rounds and bytes are properties of the protocol, so they do not depend on the hardware. If the drug, the protein, and the weights were all secret, the model would need about 435,000 comparisons per sample, and with bit decomposition the inference would take 1,727 rounds and 481\,MB per sample. Using FSS for the comparisons reduces this to 90 rounds and 41.5\,MB. Keeping only the drug secret, as the competition allows, and designing the model around it, reduces the comparisons to 12,624. With bit decomposition this still takes 1,549 rounds and 13.2\,MB per sample, and with FSS it takes 37 rounds and 0.505\,MB per sample. In total, the communication is about 950 times lower than in the first row.

\begin{table*}[t]
\centering
\caption{Communication of secure \model inference on Davis under different
settings. Rounds are per batch and do not depend on the batch size. Megabytes are
sent by each server, per sample.}
\label{tab:ablation}
\small
\begin{tabular}{@{}llrrr@{}}
\toprule
Secret inputs & Comparisons by & Comparisons per sample & Rounds & MB per sample \\
\midrule
drug, protein, weights & bit decomposition & $\approx 435{,}000$ & 1{,}727 & 481 \\
drug, protein, weights & FSS & $\approx 435{,}000$ & 90 & 41.5 \\
drug only & bit decomposition & 12{,}624 & 1{,}549 & 13.2 \\
drug only & FSS (ours) & 12{,}624 & \textbf{37} & \textbf{0.505} \\
\bottomrule
\end{tabular}
\end{table*}

Finally, the 128-bit seeds required for 128-bit security make the DCF computation 1.55 times slower than 64-bit seeds. The fused kernel more than makes up for this.

\vspace{0.5em}
\noindent\textbf{Offline Cost.} The offline phase is not timed, but it must be practical. For a batch of 512 samples, the dealer needs 78\,s to generate the material. For a batch of 1,024 samples, it needs 176\,s, 2.3 times longer, and twice the memory, for an online gain of only 5\%. We therefore use batches of 512 samples by default.

\section{Related Work}
\label{sec:related}

We review related work on DTI prediction, collaborative drug discovery, secure neural network inference, and secure inference on GPUs. For each area, we point out the gap that our work addresses.

\paragraph{Deep learning for DTI prediction.} DeepDTA~\citep{ozturk2018deepdta} predicts affinity with two convolutional towers, one for the SMILES string and one for the protein sequence. GraphDTA~\citep{nguyen2021graphdta} represents the drug as a molecular graph, MolTrans~\citep{huang2021moltrans} uses transformers, and DeepDTAGen~\citep{shah2025deepdtagen} combines affinity prediction with drug generation in one multitask model. These models are designed for accuracy in the clear. Graph convolutions and attention are among the most expensive operations to evaluate under MPC, because they multiply secret values with other secret values. We instead design \model for secure inference. It keeps every operation on the drug linear in public weights, and moves capacity to the public protein, with no loss of accuracy.

\paragraph{Collaborative drug discovery.} \citet{hie2018realizing} showed that MPC can support collaborative drug--target interaction prediction across pharmaceutical companies, using conventional machine learning models. Federated learning has also been used to train models across companies without sharing data, as in the MELLODDY project~\citep{heyndrickx2024melloddy}. These works focus on training, or on conventional models. We focus on inference with a deep model, where the drug must be protected at prediction time.

\paragraph{Secure neural network inference.} A large body of work evaluates neural networks under MPC. Early two-party systems, such as SecureML~\citep{mohassel2017secureml} and MiniONN~\citep{liu2017oblivious}, combined secret sharing with other techniques for the linear and non-linear layers. GAZELLE~\citep{juvekar2018gazelle}, Delphi~\citep{mishra2020delphi}, CrypTFlow2~\citep{rathee2020cryptflow2}, Cheetah~\citep{huang2022cheetah}, and BumbleBee~\citep{lu2025bumblebee} use homomorphic encryption for the linear layers and secure two-party protocols for the non-linear layers. Three-party systems, such as ABY3~\citep{mohassel2018aby3}, SecureNN~\citep{wagh2019securenn}, and Falcon~\citep{wagh2021falcon}, assume an honest majority among three servers to make comparisons cheaper. Function secret sharing~\citep{boyle2015function, boyle2016function} reduces a comparison to a single online round~\citep{boyle2021function}, and AriaNN~\citep{ryffel2022ariann} and LLAMA~\citep{gupta2022llama} applied it to secure inference. General frameworks, such as MP-SPDZ~\citep{keller2020mpspdz}, CrypTen~\citep{knott2021crypten}, and SecretFlow-SPU~\citep{ma2023secretflow}, make these protocols available to machine learning practitioners. Across these works, the non-linear layers dominate the cost. Several works therefore redesign the model to reduce the number of non-linear operations. CryptoNAS~\citep{ghodsi2020cryptonas} searches for architectures under a budget of ReLUs, and DeepReDuce~\citep{jha2021deepreduce} and SNL~\citep{cho2022selective} remove ReLUs from trained networks. These works target image classification and treat every input as secret. Our setting is asymmetric, with one secret input and one public input. This changes the design rules, because capacity spent on the public input is free. Moreover, most of the systems above run on CPUs.

\paragraph{Secure inference on GPUs.} A few systems run MPC on GPUs. CryptGPU~\citep{tan2021cryptgpu} showed that secret-shared computation can run on GPUs, by splitting each 64-bit value into small blocks and computing with floating-point arithmetic, an idea that we use for our linear layers. CrypTen~\citep{knott2021crypten} also supports GPU execution. Piranha~\citep{watson2022piranha} is a general platform for secret-sharing protocols on GPUs. It separates a device layer of GPU integer kernels from a protocol layer, so that the same kernels serve the two-party SecureML~\citep{mohassel2017secureml}, three-party Falcon~\citep{wagh2021falcon}, and four-party FantasticFour~\citep{dalskov2021fantastic} protocols, and it demonstrated end-to-end secure training of a realistic network, VGG, on GPUs. These systems compute comparisons with secret-sharing protocols that need many rounds or many bytes. The systems closest to ours are Orca~\citep{jawalkar2024orca}, which brought FSS to GPUs for secure training and inference, and SIGMA~\citep{gupta2024sigma}, which used GPU-accelerated FSS for secure inference of transformer models. Our DCF kernel follows the memory layout of Orca. Orca and SIGMA accelerate the cryptography for a given model, and have not been applied to DTI prediction. We combine GPU acceleration of FSS with a model designed for it, for secure DTI inference.

\section{Discussion and Limitations}
\label{sec:discussion}

\paragraph{Public protein and weights.} Our design relies on the protein and the weights being public. This matches the competition and many practical cases, but not all. If the weights were secret, every linear layer would need Beaver triples and would communicate. If the protein were also secret, the model would need about 35 times more comparisons, and the dilated convolutions of the protein tower would no longer be free. Our framework supports these cases, at a higher cost (Table~\ref{tab:ablation}).

\paragraph{Security model.} We protect against passive, or semi-honest, servers. Security against active adversaries, which may deviate from the protocol, would require additional checks, such as message authentication codes, which increase the cost. We also rely on a trusted dealer for the offline material. The dealer can be replaced by a two-party key generation protocol, at a higher offline cost.

\paragraph{Communication cost.} Our online time is now close to the time needed to send the data. The next improvements must reduce the bytes per sample. Possible directions are a smaller ring, which would halve the bytes but needs more careful truncation to avoid overflow, and fewer or cheaper comparisons. Overlapping the communication of one batch with the computation of the next would hide the remaining computation, which matters more at small batch sizes.

\paragraph{Other applications.} The asymmetry that we exploit, one secret input and one public input, appears in other applications, such as matching a private query against a public database or a public model. The key ideas of Section~\ref{sec:model} carry over to these cases.

\section{Conclusion}
\label{sec:conclusion}

The main message of this paper is the \emph{co-design of MPC-friendly models and MPC protocols}. We presented our first-place solution to Track~III of the iDASH 2026 privacy and security competition, a system for secure two-party DTI inference on GPUs.

On the model side, \model reduces the operations on the secret drug, and in particular the non-linear operations that require communication. It keeps every operation on the drug linear in public weights, and spends its capacity on the public protein, without losing accuracy. On the protocol side, we evaluate the linear layers with additive secret sharing and exact integer matrix products on the GPU, and the non-linear layers with function secret sharing and a custom CUDA kernel. Together, they took under 10\,ms per prediction on average in the competition, with secure predictions that match the plaintext model to within 0.02\%.

The ideas in this paper are not limited to DTI prediction. Many applications have one confidential input and one public input, such as a private patient record scored by a public model, or a private query matched against a public database. In such applications, a model can be designed to move its capacity to the public input, and to keep the operations on the confidential input cheap under MPC. Our GPU techniques, exact integer matrix products and a fused kernel for FSS comparisons, apply to any secret-shared inference with public weights. Our analysis also shows that, once computation is fast, the remaining cost is the network itself. Future work can therefore reduce the bytes per prediction, for example with a smaller ring or cheaper comparisons, and extend the system to active adversaries and to settings without a trusted dealer.

\vspace{0.7em}
\noindent\textbf{Acknowledgements.} This material is based upon work supported by the 
National Science Foundation 
NSF
under Grant No.~2451163 and 2523406.

\bibliographystyle{ACM-Reference-Format}
\bibliography{references}

\appendix

\begin{table*}[t]
\centering
\caption{Secure cost of \model per sample on Davis. All multiply-adds are local.
Rounds are per batch.}
\label{tab:layers}
\small
\begin{tabular}{@{}llrrr@{}}
\toprule
Layer & Output & Secret multiply-adds & Comparisons & Rounds \\
\midrule
Embedding $O E$ & $85 \times 128$ & 707{,}200 & 0 & 0 \\
Conv 1 (kernel 4) and ReLU & $16 \times 82$ & 671{,}744 & 1{,}312 & 3 \\
Conv 2 (kernel 4) and ReLU & $32 \times 79$ & 161{,}792 & 2{,}528 & 3 \\
Conv 3 (kernel 4) and ReLU & $48 \times 76$ & 466{,}944 & 3{,}648 & 3 \\
Global max pool & $48$ & 0 & 3{,}600 & 21 \\
FC 1, drug part ($48 \to 1024$), and ReLU & $1024$ & 49{,}152 & 1{,}024 & 3 \\
FC 2 ($1024 \to 512$) and ReLU & $512$ & 524{,}288 & 512 & 3 \\
Output ($512 \to 1$) & $1$ & 512 & 0 & 0 \\
Deliver the prediction & & & & 1 \\
\midrule
Total & & 2{,}581{,}632 & 12{,}624 & 37 \\
\bottomrule
\end{tabular}
\end{table*}

\section{Per-Layer Cost of \model}
\label{app:layers}

Table~\ref{tab:layers} lists the secure cost of each layer of \model on Davis, per sample. The protein tower is not listed, because it runs in plaintext.

\section{The Distributed Comparison Function}
\label{app:dcf}

Algorithms~\ref{alg:dcfgen} and~\ref{alg:dcfeval} give the key generation and evaluation of the DCF that we use, following \citet{boyle2021function}. The DCF computes $g_{\alpha,\beta}(z) = \beta \cdot \ind{z < \alpha}$ for $n$-bit inputs, with outputs in $\mathbb{Z}_{2^m}$. In our comparisons, $n = 63$, $\beta = 1$, and the outputs are bits ($m = 1$), so the additions become XORs. The PRG $G$ expands a $\lambda$-bit seed $s$ into two child seeds $s^L, s^R$, two control bits $t^L, t^R$, and two value words $v^L, v^R \in \mathbb{Z}_{2^m}$. The function $\mathrm{Convert}$ maps a seed to an element of $\mathbb{Z}_{2^m}$. We write $\alpha_i$ and $z_i$ for the $i$-th most significant bit of $\alpha$ and $z$.

Intuitively, both servers start from independent random root seeds. The control bits $t_0, t_1$ always differ on the path to $\alpha$ and are equal elsewhere. At each level, the dealer computes a correction word that a server applies whenever its control bit is 1. This makes the two servers' seeds equal as soon as the evaluation path leaves the path to $\alpha$, so that all later contributions cancel. The value correction $V_{\mathrm{CW}}$ adds $\beta$ exactly when the path leaves to the left of $\alpha$, which is when $z < \alpha$.

\begin{algorithm*}[t]
\caption{$\mathrm{Gen}(\alpha, \beta)$, run by the dealer offline.}
\label{alg:dcfgen}
\begin{algorithmic}[1]
\State Sample random $\lambda$-bit seeds $s_0, s_1$; \ set $t_0 \gets 0$, $t_1 \gets 1$, $V_\alpha \gets 0$
\State Keep the root seeds $s_0^{(0)} \gets s_0$, $s_1^{(0)} \gets s_1$
\For{$i = 1$ to $n$}
  \State $(s_b^L, v_b^L, t_b^L, s_b^R, v_b^R, t_b^R) \gets G(s_b)$ for $b \in \{0, 1\}$
  \If{$\alpha_i = 0$} $\mathrm{Keep} \gets L$, $\mathrm{Lose} \gets R$ \Else\ $\mathrm{Keep} \gets R$, $\mathrm{Lose} \gets L$ \EndIf
  \State $s_{\mathrm{CW}} \gets s_0^{\mathrm{Lose}} \oplus s_1^{\mathrm{Lose}}$
  \State $V_{\mathrm{CW}} \gets (-1)^{t_1} \cdot \big(v_1^{\mathrm{Lose}} - v_0^{\mathrm{Lose}} - V_\alpha + \alpha_i \cdot \beta\big)$
  \State $V_\alpha \gets V_\alpha - v_1^{\mathrm{Keep}} + v_0^{\mathrm{Keep}} + (-1)^{t_1} \cdot V_{\mathrm{CW}}$
  \State $t_{\mathrm{CW}}^L \gets t_0^L \oplus t_1^L \oplus \alpha_i \oplus 1$; \quad $t_{\mathrm{CW}}^R \gets t_0^R \oplus t_1^R \oplus \alpha_i$
  \State $\mathrm{CW}^{(i)} \gets (s_{\mathrm{CW}}, V_{\mathrm{CW}}, t_{\mathrm{CW}}^L, t_{\mathrm{CW}}^R)$
  \For{$b \in \{0, 1\}$}
    \State $s_b \gets s_b^{\mathrm{Keep}} \oplus t_b \cdot s_{\mathrm{CW}}$; \quad $t_b \gets t_b^{\mathrm{Keep}} \oplus t_b \cdot t_{\mathrm{CW}}^{\mathrm{Keep}}$
  \EndFor
\EndFor
\State $\mathrm{CW}^{(n+1)} \gets (-1)^{t_1} \cdot \big(\mathrm{Convert}(s_1) - \mathrm{Convert}(s_0) - V_\alpha\big)$
\State \Return $k_b = \big(s_b^{(0)}, \mathrm{CW}^{(1)}, \ldots, \mathrm{CW}^{(n+1)}\big)$ for $b \in \{0, 1\}$
\end{algorithmic}
\end{algorithm*}

\begin{algorithm*}[t]
\caption{$\mathrm{Eval}(b, k_b, z)$, run by server $P_b$ online.}
\label{alg:dcfeval}
\begin{algorithmic}[1]
\State Parse $k_b = \big(s, \mathrm{CW}^{(1)}, \ldots, \mathrm{CW}^{(n+1)}\big)$; \ set $t \gets b$, $V \gets 0$
\For{$i = 1$ to $n$}
  \State Parse $\mathrm{CW}^{(i)} = (s_{\mathrm{CW}}, V_{\mathrm{CW}}, t_{\mathrm{CW}}^L, t_{\mathrm{CW}}^R)$
  \State $(s^L, v^L, t^L, s^R, v^R, t^R) \gets G(s)$
  \If{$t = 1$}
    \State $s^L \gets s^L \oplus s_{\mathrm{CW}}$; \ $s^R \gets s^R \oplus s_{\mathrm{CW}}$; \ $t^L \gets t^L \oplus t_{\mathrm{CW}}^L$; \ $t^R \gets t^R \oplus t_{\mathrm{CW}}^R$
  \EndIf
  \If{$z_i = 0$}
    \State $V \gets V + (-1)^b \cdot (v^L + t \cdot V_{\mathrm{CW}})$; \ $s \gets s^L$; \ $t \gets t^L$
  \Else
    \State $V \gets V + (-1)^b \cdot (v^R + t \cdot V_{\mathrm{CW}})$; \ $s \gets s^R$; \ $t \gets t^R$
  \EndIf
\EndFor
\State \Return $V + (-1)^b \cdot \big(\mathrm{Convert}(s) + t \cdot \mathrm{CW}^{(n+1)}\big)$
\end{algorithmic}
\end{algorithm*}

Each key holds one $\lambda$-bit root seed and, per level, a $\lambda$-bit seed correction, a value correction, and two control bits. With $n = 63$ and $\lambda = 128$, a key holds about one kilobyte of correction words. Since the correction words are the same in both keys, the dealer can store them once per comparison. An evaluation makes $n$ calls to $G$, each of which costs two AES evaluations in our implementation (Equation~\eqref{eq:mmo}).

\section{Exactness of the Limb Matrix Product}
\label{app:limb}

Let $A$ and $B$ be matrices over $\mathbb{Z}_{2^{64}}$ with inner dimension $K$, and write $A = \sum_{i=0}^{3} A_i 2^{16 i}$ and $B = \sum_{j=0}^{3} B_j 2^{16 j}$, with all entries of the limb matrices in $[0, 2^{16})$. Then
\begin{equation}
A B \equiv \sum_{i + j < 4} (A_i B_j) \, 2^{16 (i + j)} \pmod{2^{64}} .
\end{equation}
Each entry of $A_i B_j$ is a sum of $K$ products, each at most $(2^{16} - 1)^2 < 2^{32}$, so it is an integer smaller than $K \cdot 2^{32}$. A 64-bit floating-point number represents every integer up to $2^{53}$ exactly, and the floating-point matrix product computes every partial sum exactly as long as no partial sum exceeds $2^{53}$. This holds when $K \cdot 2^{32} \le 2^{53}$, that is, $K \le 2^{21}$. The largest inner dimension in \model is 1,024, the hidden width of the prediction head, so all products are exact. We convert each exact result back to a 64-bit integer, shift it by $16 (i + j)$ bits, and add the ten terms modulo $2^{64}$. The int8 variant works the same way with eight 8-bit limbs. Only the products with $i + j < 8$ survive, and each is computed with 8-bit integer inputs and 32-bit integer accumulation on the tensor cores.

\section{Local Truncation}
\label{app:trunc}

Let $\share{x} = (x_0, x_1)$ hold a fixed-point value with $2f$ fractional bits. Following \citet{mohassel2017secureml}, the servers compute
\begin{equation}
x_0' = \lfloor x_0 / 2^{f} \rfloor, \qquad
x_1' = 2^{\ell} - \lfloor (2^{\ell} - x_1) / 2^{f} \rfloor \bmod 2^{\ell} .
\end{equation}
If $|x| < 2^{\ell_x}$ for some $\ell_x < \ell - 1$, then $x_0' + x_1'$ equals $\lfloor x / 2^{f} \rfloor$ up to an error of one in the least significant bit, except with probability at most $2^{\ell_x + 1 - \ell}$. In that rare case, the result is wrong by a large amount, because the two shares wrap around the ring in different ways. With $\ell = 64$, our values stay far below $2^{63}$, so the probability is small.

\end{document}